\documentclass[11pt]{article}
\usepackage[left=20mm, right=20mm, top=15mm, bottom=15mm, includehead=true, includefoot=true]{geometry}

\usepackage[colorlinks=true,allcolors=black,citecolor=blue, linkcolor=blue]{hyperref}
\usepackage{listings}
\usepackage{color}
\usepackage{graphicx}
\usepackage{float}
\usepackage{graphicx}
\usepackage{natbib} 
\usepackage{url}
\usepackage{authblk} 
\usepackage{todonotes}
\usepackage{float}
\usepackage{listings}
\usepackage{rotating}
\usepackage{multirow}
\usepackage{multicol}
\usepackage{booktabs}
\usepackage{enumitem}
\usepackage{longtable}
\usepackage{tablefootnote}
\usepackage[parfill]{parskip} 
\usepackage{rotating}
\usepackage{array,etoolbox}
\usepackage{setspace}
\usepackage{lineno}
\usepackage{tikz}
\usepackage[tikz]{mdframed}

\definecolor{light-gray}{gray}{0.95}
\newmdenv[%
outerlinewidth=2,%
roundcorner=10pt,%
linecolor=gray,%
backgroundcolor=light-gray,%
]{myframe}

\usepackage{tikz}

\usepackage{helvet}

\preto\tabular{\setcounter{magicrownumbers}{0}}
\newcounter{magicrownumbers}

\usepackage{sectsty}
\sectionfont{\normalsize}
\subsectionfont{\normalsize}
\subsubsectionfont{\normalsize}
\paragraphfont{\normalsize}

\newcommand{\ie}{i.e.}
\newcommand{\eg}{e.g.}

\newcommand{\Reffig}[1]{Figure~\ref{#1}}
\newcommand{\Refsec}[1]{Section~\ref{#1}}
\newcommand{\Reftab}[1]{Table~\ref{#1}}

\definecolor{gray}{rgb}{0.4,0.4,0.4}
\definecolor{darkblue}{rgb}{0.0,0.0,0.6}
\definecolor{cyan}{rgb}{0.0,0.6,0.6}
\definecolor{mauve}{rgb}{0.58,0,0.82}

\lstdefinelanguage{XML}
{
  frame=tb,
  aboveskip=3mm,
  belowskip=3mm,
  breaklines=true,
  breakatwhitespace=true,
  morestring=[b]",
  morestring=[s]{>}{<},
  morecomment=[s]{<?}{?>},
  stringstyle=\color{darkblue},
  identifierstyle=\color{darkblue},
  keywordstyle=\color{mauve},
  morekeywords={xmlns,version,type,name,codeSpace}
}

\title{Reference study of IFC software support: the GeoBIM benchmark 2019 --- Part I}

\author[1]{Francesca Noardo}
\author[1]{Thomas Krijnen}
\author[1]{Ken Arroyo Ohori}
\author[2,3]{Filip Biljecki}
\author[4]{Claire Ellul}
\author[5]{Lars Harrie}

\author[5]{Helen Eriksson}
\author[6]{Lorenzo Polia}
\author[1]{Nebras Salheb}
\author[7]{Helga Tauscher}
\author[1]{Jordi van Liempt}

\author[8]{Hendrik Goerne}
\author[9]{Dean Hintz}
\author[7]{Tim Kaiser}
\author[10]{Cristina Leoni}
\author[11]{Artur Warchoł}

\author[1]{Jantien Stoter}

\affil[1]{3D Geoinformation, Delft University of Technology, Delft, The Netherlands --- f.noardo@tudelft.nl, t.f.krijnen@tudelft.nl, k.ohori@tudelft.nl, n.salheb@hotmail.com, j.n.h.vanliempt@tudelft.nl, j.e.stoter@tudelft.nl}

\affil[2]{Department of Architecture, National University of Singapore, Singapore --- filip@nus.edu.sg}
\affil[3]{Department of Real Estate, National University of Singapore, Singapore}

\affil[4]{Department of Civil, Environmental and Geomatic Engineering, University College London, London, UK --- c.ellul@ucl.ac.uk}

\affil[5]{Department of Physical Geography and Ecosystem Science, Lund University, Sweden --- (lars.harrie, helen.eriksson)@nateko.lu.se}

\affil[6]{Architect --- lorenzo.polia@libero.it}

\affil[7]{Faculty of Spatial Information, Dresden University of Applied Sciences, Dresden, Germany --- (helga.tauscher, tim.kaiser)@htw-dresden.de}

\affil[8]{GMX, Germany --- HendrikGoerne@gmx.de}

\affil[9]{Safe Software, Surrey, Canada --- dean.hintz@safe.com}

\affil[10]{Department of Civil, Constructional and Environmental Engineering, Sapienza Univerisity of Rome, Rome --- cristina.leoni@uniroma1.it}

\affil[11]{Institute of Technical Engineering, PWSTE Bronisław Markiewicz State University of Technology and Economics in Jarosław, Poland / ProGea 4D sp. z o.o. --- artur.warchol@pwste.edu.pl}

\begin{document}
	
	\maketitle

\begin{myframe}
	
	This is the author's version of the work. 
	
	It is posted here only for personal use, not for redistribution and not for commercial use.
	
	The definitive version is published in the journal \emph{Transactions in GIS}.
	\\	
	\\
	Noardo, F., Krijnen, T., Arroyo Ohori, K., Biljecki, F., Ellul, C., Harrie, L., Eriksson, H., Polia, L., Salheb, N., Tauscher, H., van Liempt, J., Goerne, H., Hintz, D., Kaiser, T., Leoni, C., Warchol, A., Stoter, J. (2021). Reference study of IFC software support: the GeoBIM benchmark 2019 – Part I. \emph{Transactions in GIS}.
	\\\textsc{doi}: \url{https://doi.org/10.1111/tgis.12709}
	\\
	\\
	Full details of the project: \url{https://3d.bk.tudelft.nl/projects/geobim-benchmark/}
	
\end{myframe}

\doublespacing


 \begin{abstract}
IFC, the buildingSMART open standard for BIM, is underused with respect to its promising potential, since, according to the experience of practitioners and researchers working with BIM, issues in the standard's implementation and use prevent its effective use.
Nevertheless, a systematic investigation of these issues has never been performed, and there is thus insufficient evidence for tackling the problems.
The GeoBIM benchmark project is aimed at finding such evidence by involving external volunteers, reporting on tools behaviour about relevant aspects (geometry, semantics, georeferencing, functionalities), analysed and described in this paper.
Interestingly, different IFC software with the same standardised datasets yield inconsistent results, with few detectable common patterns, and significant issues are found in their support of the standard, probably due to the very high complexity of the standard data model.
This paper is in tandem with Part II, describing the results of the benchmark related to CityGML, the counterpart of IFC within geoinformation\@.

 $ $ \\ {\bf KEYWORDS:} Industry Foundation Classes, Building Information Models, open standards, software support, interoperability, GeoBIM\@.

 \end{abstract}


\section{Introduction}\label{sec:intro}

In the Architecture, Engineering and Construction (AEC) fields, as well as in further disciplines, interoperability is of increasing importance, in order to enable re-use and exchange of data and information, including in the strictly related asset and facility management field.
Furthermore, it is essential for reciprocal integration of data having different nature.
One of the current research topics is, for example, the integration of building information models (BIMs) with 3D city models, for supporting several use cases, \eg\ building permits issuing, 3D cadastre, complex assets and facility management.
International standards are conceived as a solution for fostering such interoperability.
The most popular standard data models, considered for obtaining such integration, are usually the OGC CityGML for 3D city models and buildingSMART Industry Foundation Classes (IFC) for BIMs.

However, although a part of the building designers and professionals, supporting openBIM, are IFC enthusiasts, many others working with BIMs still seldom use IFC as a first choice for exchanging their models, and only export them to IFC when required explicitly by law, when trying to integrate with software from different vendors, or when no other solutions are available, as emerged from several personal communications.
This often happens because they are aware of some limitations in the IFC format (as for example documented more than ten years ago by \citet{pazlar2008interoperability}, or in \citet{barbini2019bim}), although they might acknowledge its potential as an open exchange format.

In practice, limited support for IFC might not been considered as a major issue, since some very widespread software, such as the \textit{Autodesk Revit} format, is often used as  \textit{de facto} standard for BIMs exchange among designers, as well as for integrating the BIM model with further systems (\citet{baik2015integration,kensek2014integration,petersen2018integration,Lv2018/09,kamari2018bim,papadopoulos2017evaluation,lamartina2019structural,aksamija2018methods,farid2019integration}
are only few examples), while the actual use of BIMs by authorities, besides visualization, is still limited, for the moment.
This solves the most immediate interoperability issues, even though the \textit{Revit} format is a proprietary binary format, which hinders re-usability in an open way and across time.

A specific study of the limitations of IFC was part of the GeoBIM benchmark project\footnote{\url{https://3d.bk.tudelft.nl/projects/geobim-benchmark/}} (see Section~\ref{sec:intro2}) and is reported in this paper.
Within the project, the approach to the study of the support for the two standards involved in the GeoBIM integration (IFC and CityGML) was conceived in parallel, also with the aim of understanding if one of the two offered more effective solutions that could be possibly borrowed by the other one in future developments.
However, the final outcomes of the two different tasks are very specific for each standard and deserve to be presented and discussed separately, considering the specificities of each case.
For these reasons, this paper, which focuses on the results about the benchmark Task 1 --- support for IFC, is written in tandem with \citet{noardo2020referenceII}, which describes Task 3, covering the support for CityGML\@.
In order to allow each paper to be read on its own, the two papers share some information (\ie\ Section~\ref{sec:intro2} explaining the general context and motivation of the study; Section~\ref{sec:metbench} covering the initial part of the methodology about the entire GeoBIM benchmark set-up, and Section~\ref{sec:mettask13} concerning some similarities in the methodology).
One further paper explores the parts of the project more directly related to the subject of integration itself, namely, conversion procedures and useful tools to georeference IFC models \citep{noardo2020tools}. 

\section{The GeoBIM needs and the concept of this study}\label{sec:intro2}

Two increasingly developed, studied and used 3D information systems have revealed their potential in the related fields:

\begin{itemize}
	\item \textbf{3D city models}, which are used to represent city objects in three dimensions and advance previous 2D maps and other cartographic products, in order to support city analysis and management, city planning, navigation, and so on (\eg\ 
	\citet{biljecki2015applications,Kumar17,egusquiza2018energy,Jakubiec13,Liang14,Bartie10,Peters15,Nguyen12});
	\item	\textbf{Building Information Models (BIM)}, which are used in the architecture, engineering and construction fields (AEC) to design and manage buildings, infrastructure and other construction works, and which also have features useful to project and asset management (\eg\ 
	\citet{petri2017optimizing,haddock2018building,azhar2011building}%
).
\end{itemize}

Several international standards exist to rule the representation of the built environment in a shared way, to foster interoperability and cross border analysis and, consequently, actions, or to reuse tools, analysis methods and data themselves for research and, possibly, government.
Some example of international standards are: the European Directive for an infrastructure for spatial information in Europe (INSPIRE)\footnote{\url{https://inspire.ec.europa.eu}}, aimed at the representation of cross border pieces of land in Europe, for common environmental analysis; the Land and Infrastructure standard (LandInfra)\footnote{\url{https://www.ogc.org/standards/landinfra}}, by the Open Geospatial Consortium (OGC), aimed at land and civil engineering infrastructure facilities representation; and the green building data model (gbXML)\footnote{\url{https://www.gbxml.org}}, aimed at the representation of buildings for energy analysis.

Nonetheless, the two dominant reference open standards for those two models are CityGML\footnote{\url{citygmlwiki.org}}, by the OGC, focusing on urban-scale representation of the built environment, and the Industry Foundation Classes~\citep{ISO16739:2013}\footnote{\url{ https://technical.buildingsmart.org/standards/ifc/}}, by buildingSMART, aimed at the very detailed representation of buildings and other construction works for design and construction objectives, first, but also intended to enable project management throughout the process, and asset and facility management in a following phase. Those standards are both intended to be very comprehensive and are therefore very wide and articulated.
They both use complex data models allowing for a wide variety of models using object-oriented representations, even if that comes at a cost of slower and more inconsistent implementations.

Due to the overlapping interests in both fields (meeting in the building-level representation), increasing attention is being paid to 3D city model-BIM integration (GeoBIM), where the exchange of information between geospatial (3D city models) and BIM sources enables the reciprocal enrichment of the two kinds of information with advantages for both fields, \eg\ automatic updates of 3D city models with high-level-of-detail features, automatic representation of BIM in their context, automated tests of the design, and so on~\citep{liu2017state,fosu2015integration,aleksandrov2019systems,kumar2019landinfra,niu2019logistics,Noardo19b,Arroyo-Ohori18a,kang2015study,Stouffs:2018kg,Lim:2019vh}.

The GeoBIM subject can be divided into several sub-issues.

\begin{enumerate}
	\item The harmonization of data themselves, which have to concretely fit together, with similar (or harmonizable) features (\eg\ accuracy, kind of geometry, amount of detail, kind of semantics, georeferencing).
	
	\item Interoperability, which is a fundamental key in the integration.
	It is important to note here, that before enabling the interoperability among different formats (\eg\ GIS formats and BIM formats), which is the theme of point three below, the interoperability GIS-to-GIS and BIM-to-BIM itself is essential. That means that the formats of data have to be understood and correctly interpreted uniquely by both any person and any supporting software. Moreover, an interoperable dataset is supposed to remain altogether unchanged when going through a potentially infinite number of imports and exports by software tools, possibly converting it to their specific native formats and exporting it back.
	
	\item The effective conversion among different formats, \ie\ transforming one dataset in a (standardised) format to another one in compliance with the end format specifications and features.
	
	\item The procedures employing 3D city models and the ones based on BIM should be changed in order to obtain better advantages by the use of both, integrated, since those systems enable processes which are usually more complex than just the simple representations.
\end{enumerate}

The many challenges implied by the points above are still far from being solved, and one of the essential initial steps is actually to outline such challenges more sharply.

In particular, the second point (interoperability and involved standards) is often considered to be solved by standardization organizations.
It is desirable to rely on open standards for this, because the well-documented specifications of open standards would enable longer-term support, as well as their genericity with respect to different software vendors, as opposed to closed point-to-point solutions that merely connect one proprietary system to another (and might be discontinued or stop working at any moment).
However, our previous experiences suggest that the support for open standards in software often include shortcomings.

The researchers promoting this study (as users of data, advocates of open standards and developers of tools adopting such standards) have noticed, over their research and work activities, how the use of those standards in data and their implementations in software were not always straightforward and not completely consistent with the standard specifications either.
Many tools, when managing standardized data, do not adequately support features or functionalities as they do when the data is held in the native formats of the software.
In addition, software tools have limitations with respect to the potential representation (geometry, semantics, georeferencing) of data structured following these standards, or can generate errors and erroneous representations by misinterpreting them.

The standards themselves are partly at fault here, since they often leave some details undefined, with a high degree of freedom and various possible interpretations.
They allow high complexity in the organization and storage of the objects, which does not work effectively towards universal understanding, unique implementations and consistent modelling of data. This is probably due to the fact that such standards often originate as amalgamations of existing mechanisms and compromises between the various stakeholders involved.
These experiences have been informally confirmed through exchanges within the scientific community and especially with the world of practitioners, who are supposed to work with (and have the most to gain from) those standardized data models and formats.
However, more formal evidence on the state of implementation of these open standards and what problems could be connected to the standard themselves have not been compiled so far.

For this reason, the GeoBIM benchmark project\footnote{\url{https://3d.bk.tudelft.nl/projects/geobim-benchmark/}}\(^,\)\footnote{\url{https://www.isprs.org/society/si/SI-2019/TC4-Noardo\_et\_al\_WG-IV-2-final\_report.pdf}} was proposed and funded in 2019 by the International Society for Photogrammetry and Remote Sensing (ISPRS)\footnote{\url{https://www.isprs.org}} and the European association for Spatial Data Research (EuroSDR)\footnote{\url{http://www.eurosdr.net}}.
The aim of the benchmark was to get a better picture of the state of software support for the two open standards (IFC and CityGML) and the conversions between them, in order to formulate recommendations for further development of the standards and the software that implements them.
In addition, we tested two known major technical issues related to GeoBIM integration and which are known to be solved only partially in practice: the ability of tools and methods to georeference IFC and the conversion procedures between IFC and CityGML\@ \citep{noardo2020tools}.

The relevant outcomes regarding the support of software for buildingSMART IFC standard are the subject of this paper.

\subsection{Industry Foundation Classes and knotty points}\label{sec:IFC}

The buildingSMART Industry Foundation Classes (IFC) standard\footnote{\url{https://technical.buildingsmart.org/standards/ifc/}} is an open standard data model for Building Information Modelling (BIM) to be shared and exchanged through software applications, domains and use cases, within the Architecture Engineering and Construction (AEC) and Facility Management (FM) fields.
It includes classes for describing both physical and abstract concepts (\eg\ cost, schedule, etc.) concerning AEC-FM for buildings, mainly (recent versions are extending it for including infrastructures and other kinds of constructions)\footnote{\url{https://technical.buildingsmart.org/standards/ifc/ifc-schema-specifications/}}.
It has also been adapted as the ISO 16739 international standard~\citep{ISO16739:2013}.

The standard includes relevant constructs for a wide variety of disciplines, use cases and processes associated to the construction domain, most prominently the semantic description and geometric representation of typical construction elements and their relationships.

The IFC is structured in a deeply hierarchical data model, furthermore organized in several and deep and complex meronymic (part-of) trees too.
The spatial composition (Site/Building/Storey/Space/Zone) is one more kind of aggregation, different from the element (meronymical) composition one (\eg\ a stair and the assembled elements in it).
Moreover, nesting is a slightly different kind of element composition, representing the products which are specifically designed as complementary by manufacturers.
Finally, subtraction relationships are also part of the IFC model, representing openings by means of the voiding mechanism.
A great number of further relationships are added to this complexity (\eg\ to associate materials, geometric representation or other property information and so on).
 
An additional complexity to the semantic model is given by the possibility to store the same kind of object by means of several entities. For example the layers within a compound wall object can be represented by means of an associated \texttt{IfcMaterialLayerSet}, but also as a more generic decomposition where every wall layer is modelled as a distinct \texttt{IfcBuildingElementPart}.

A great number of attributes can be associated with entities (and inherited by the parent-ones), both directly or through property sets.

All this semantic complexity is intended to represent faithfully the buildings as functional to the standard designed scope.
However, the implementation and use of such theoretically precise model is difficult and can result in inaccuracies or under-use of it, besides hindering interoperability for leaving too high freedom in filling the information in and in choosing the kind of representation to be used.

Additional terms which can be used in IFC, are defined within the buildingSMART Data Dictionaries (bsDD) and are modelled according to the International Framework for Dictionaries (IFD)\footnote{\url{http://bsdd.buildingsmart.org}}.
It is based on the standard ISO 12006--3.

The IFC current versions are: IFC2x3, which was released in 2005 (with the latest corrigendum in 2007) and the IFC4.1 from 2018.
At the time of writing, the most implemented and used version is still (by far) IFC2x3.
For this reason, both versions were considered in this study.

A third part of the standards is the Information Delivery Manual (IDM), which defines the workflow and the information exchange specifications and requirements in the processes involved in the construction life cycle.

In an IDM, a set of Model View Definitions (MVD) can be defined for identifying the portion of the IFC model which is needed for a particular information exchange in the IDM to be fulfilled. This can define a use-case oriented part of the wide IFC model, to be implemented in software.

Interesting to note is that this mechanism is in a way opposite of CityGML Application Domain Extensions (ADEs)~\citep{Biljecki2018}. In an IFC MVD a subset of definitions is selected from the monolithic schema, instead of an addition. On the other hand, for property set definitions, the usage of MVD and ADE are analogous, as both nominate additional attributes.

IFC derives many aspects from \citet{ISO10303:2014}, informally known as STEP\@. The majority of geometry definitions are derived from ISO 10303--42 and the typical exchange formats are based on STEP Physical File (SPF, ISO 10303--21) and an XML encoding (ISO 10303--28).

Parametric modelling is usually employed in BIM and IFC, which makes it possible to encode many kinds of geometries.
This includes Boolean operations and complex sweeps, for example the sweep of an arbitrary profile along a curve while constraining the normal vector.
Also explicit geometries are supported in the form of Boundary Representations and (added in IFC4) efficient support for triangulated meshes.
The implementation of the former type of geometry is that supporting the full stack of geometric procedures in IFC is a major implementation effort and due to implementation choices can sometimes lead to different results in importing applications.
The complexity can, therefore, have consequences on interoperability and the way different pieces of software read and re-export the same geometry.

\subsubsection{IFC geometries}\label{sec:introIFCgeom}

\begin{figure}[h]
	\centering
    \includegraphics[width=1\linewidth]{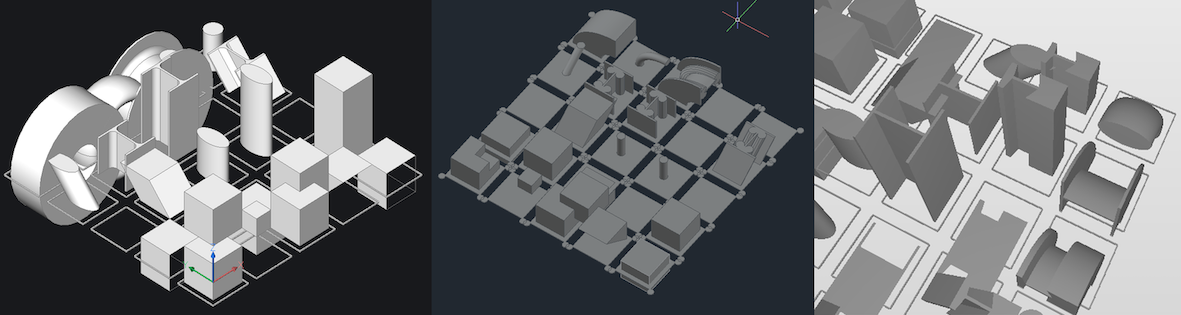}
    \caption{Example of unusual behaviour of the geometries in some of the tested software.}%
    \label{fig:weirdifcgeom}
\end{figure}

The particular issues described in this section guided the methodological choice of providing a specific set of IFC geometries among the benchmark datasets, in order to point out specific related behaviour (\Reffig{fig:weirdifcgeom} is an example of uncontrolled geometries transformations within some of the tested software).

IFC is a complex standard and the fidelity of both importers and exporters is still evolving.
Implementations are based on varying paradigms and geometric modelling principles: Boundary Representation, Polyhedra or Triangular meshes.
Some implementations use so called toolkits that offer complete support for STEP and EXPRESS (the modelling language in which the IFC schema is specified; with an embedded complex Turing-complete constraint language to specify WHERE rules and FUNCTIONS).
Other implementations use custom in-house developments and, for that reason, typically have less support for the automatic validation of IFC models based on the WHERE rules.
The IFC schema is evolving as well\footnote{as seen in repositories such as \url{https://github.com/buildingSMART/NextGen-IFC}} mostly aiming at the improvement of reliability, reuse of industry-standard best-practices and the reduction of implementation effort.

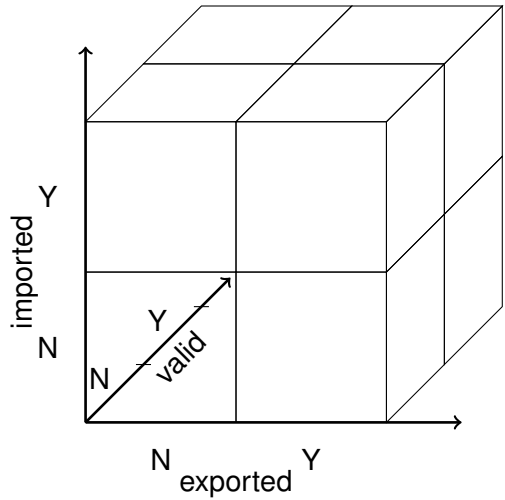
\begin{figure}
\centering
	\begin{tikzpicture}
	
	\coordinate (n000) at (0,0,0);
	\coordinate (n010) at (0,2,0);
	\coordinate (n011) at (0,2,2);
	\coordinate (n001) at (0,0,2);
	\coordinate (n100) at (2,0,0);
	\coordinate (n110) at (2,2,0);
	\coordinate (n111) at (2,2,2);
	\coordinate (n101) at (2,0,2);
	\coordinate (n200) at (4,0,0);
	\coordinate (n210) at (4,2,0);
	\coordinate (n211) at (4,2,2);
	\coordinate (n201) at (4,0,2);
	
	\coordinate (n001) at (0,0,0+2);
	\coordinate (n011) at (0,2,0+2);
	\coordinate (n012) at (0,2,2+2);
	\coordinate (n002) at (0,0,2+2);
	\coordinate (n101) at (2,0,0+2);
	\coordinate (n111) at (2,2,0+2);
	\coordinate (n112) at (2,2,2+2);
	\coordinate (n102) at (2,0,2+2);
	\coordinate (n201) at (4,0,0+2);
	\coordinate (n211) at (4,2,0+2);
	\coordinate (n212) at (4,2,2+2);
	\coordinate (n202) at (4,0,2+2);
	
	\coordinate (n010) at (0,0+2,0);
	\coordinate (n020) at (0,2+2,0);
	\coordinate (n021) at (0,2+2,2);
	\coordinate (n011) at (0,0+2,2);
	\coordinate (n110) at (2,0+2,0);
	\coordinate (n120) at (2,2+2,0);
	\coordinate (n121) at (2,2+2,2);
	\coordinate (n111) at (2,0+2,2);
	\coordinate (n210) at (4,0+2,0);
	\coordinate (n220) at (4,2+2,0);
	\coordinate (n221) at (4,2+2,2);
	\coordinate (n211) at (4,0+2,2);
	
	\coordinate (n011) at (0,0+2,0+2);
	\coordinate (n021) at (0,2+2,0+2);
	\coordinate (n022) at (0,2+2,2+2);
	\coordinate (n012) at (0,0+2,2+2);
	\coordinate (n111) at (2,0+2,0+2);
	\coordinate (n121) at (2,2+2,0+2);
	\coordinate (n122) at (2,2+2,2+2);
	\coordinate (n112) at (2,0+2,2+2);
	\coordinate (n211) at (4,0+2,0+2);
	\coordinate (n221) at (4,2+2,0+2);
	\coordinate (n222) at (4,2+2,2+2);
	\coordinate (n212) at (4,0+2,2+2);
	
	\draw[black] (n200) -- (n210) -- (n211) -- (n201) -- cycle;
	\draw[black] (n002) -- (n012) -- (n112) -- (n102) -- cycle;
	\draw[black] (n201) -- (n211) -- (n212) -- (n202) -- cycle;
	\draw[black] (n102) -- (n202) -- (n212) -- (n112) -- cycle;
	\draw[black] (n020) -- (n021) -- (n121) -- (n120) -- cycle;
	\draw[black] (n210) -- (n220) -- (n221) -- (n211) -- cycle;
	\draw[black] (n121) -- (n221) -- (n220) -- (n120) -- cycle;
	\draw[black] (n012) -- (n022) -- (n122) -- (n112) -- cycle;
	\draw[black] (n021) -- (n022) -- (n122) -- (n121) -- cycle;
	\draw[black] (n211) -- (n221) -- (n222) -- (n212) -- cycle;
	\draw[black] (n112) -- (n212) -- (n222) -- (n122) -- cycle;
	\draw[black] (n122) -- (n222) -- (n221) -- (n121) -- cycle;
	
	\node at (1,-0.5,4) {N};
	\node at (3,-0.5,4) {Y};
	
	\node at (-0.5,1,4) {N};
	\node at (-0.5,3,4) {Y};
	
	\node at (-0.2,0.2,3) {N};
	\node at (-0.2,0.2,1) {Y};
	
	\draw[black, arrows=->, line width=1pt] (0,0,4) -- (5,0,4);
	\draw[black, arrows=->, line width=1pt] (0,0,4) -- (0,0,-1);
	\draw[black] (-0.1,0,2) -- (0.1,0,2);
	\draw[black] (-0.1,0,0) -- (0.1,0,0);
	\draw[black, arrows=->, line width=1pt] (0,0,4) -- (0,5,4);
	
	\path (1,-0.5,4) -- (3,-0.5,4) node [midway, below, sloped] {exported};
	\path (-0.5,1,4) -- (-0.5,3,4) node [midway, above, sloped] {imported};
	\path (0,0,2) -- (0,0,0.5) node [midway, below, sloped] {valid};
	
	\end{tikzpicture}%
	\caption{Three-dimensional matrix representing the IFC geometries conditions: can be exported / can be imported / is valid according to the IFC standard.}%
	\label{fig:ifc3val}
\end{figure}

Given these considerations, the permutations of IFC definitions (some of which assessed by this set of analytical geometries) span a three-dimensional matrix along the axes (\Reffig{fig:ifc3val}) that specify:

\begin{enumerate}
	\item whether the situation is known to be exported by authoring tools
	\item whether the situation is handled successfully by importing applications and
	\item whether that situation is valid according to the IFC schema.
\end{enumerate}

As such we can summarize this state in a tuple of three elements.
For example, for a situation that is exported by authoring tools, successfully imported, but not valid according to the schema, the tuple would be $ < Y, Y, N > $.

From these states we want to highlight three that are meaningful:

\begin{itemize}
	\item $ < N, Y/N, Y/N > $ is a situation that is never exported and can hence be removed from the schema to reduce the implementation effort of a fully compliant application.
	\item $ < Y, Y, N > $ are situations that are successfully imported, but not valid according to the schema, schema constraints may be loosened to actually allow such situations.
	\item $ < Y, N, Y/N > $ is the biggest problem for practitioners, \ie\ geometries that are exported, but not successfully imported (whether valid or not).
\end{itemize}

An additional concern is that while validity according to the schema is fairly constant and universal, import success depends on the application used, and with many different tools used by practitioners, the diverging results of importers may cause misconceptions and disputes.
Also note that validity is not as universal as it might seem at first glance.
The IFC schema defines a point equality tolerance (\textit{IfcGeometricRepresentationContext.Precision}) that defines ``the tolerance under which two given points are still assumed to be identical''.
How this is handled in geometry kernels is implementation dependent.
Additionally there are WHERE rules that specify that an extrusion length needs to be positive (\textit{IfcExtrudedAreaSolid.Depth} is of type \textit{IfcPositiveLengthMeasure} which has a rule \texttt{WR1 : SELF > 0.}) but at the same time EXPRESS does not specify restrictions on the floating point number type, so an infinitesimal value can be supplied that is greater than zero, but when mapped to the number type in the importing application cannot be distinguished from it.
Or an extrusion depth can be supplied that is below the model tolerance and therefore may fail in applications that adhere to this tolerance.

A significant set of geometries with a variation of such three features was added as part of the tested IFC datasets, as described in Section~\ref{sec:IFCgeometries}.

%

\subsubsection{IFC georeferencing}\label{sec:IFCgeoref}

Proper georeferencing of an IFC file facilitates a link between the local Cartesian coordinate system used by IFC, with their corresponding real-world coordinate reference system (CRS), and thus to place the model of a single building or construction within its context and environment. There are several options to store georeferencing information in IFC, with varying level of detail. These options range from basic address information over the specification of the geolocation of a reference point to the definition of an offset between the project coordinate system and the global origin of a coordinate reference system and the corresponding rotation of the XY-Plane. Some of the georeferencing options are described and classified into Levels of Georeferencing (LoGeoRefxx) by \citet{clemen2019level} (\Reftab{tab:LoGeoRefs}). This classification scheme is not officially defined in the IFC standard but can help practitioners to assess which georeferencing information is available in the IFC file. Common for the presented options are that coordinates for a reference point (which is usually the origin of the local Cartesian system where the model is designed) are stored and sometimes complemented with a direction of the axis in the local system.

\begin{table}[H]
	\centering
	\small
	\begin{tabular}{|m{2cm}|m{4cm}|m{10cm}|}
		\hline
		\textbf{LoGeoRef} & \textbf{Supported CRS} & \textbf{Storing entities} \\ \hline
		\textit{LoGeoRef10} & No CRS, approximate location by means of the address. & \textit{IfcPostalAddress} referenced by either \textit{IfcSite} or \textit{IfcBuilding}. \\ \hline
		\textit{LoGeoRef20}& WGS84 EPSG:4326 & Attributes \textit{RefLatitude, RefLongitude, RefElevation} within \textit{IfcSite} \\ \hline
		\textit{LoGeoRef30} & Any Cartesian CRS, including projected coordinates (CRS not specified in the file) & \textit{IfcCartesianPoint} referenced within \textit{IfcSite} (defining the projected coordinates of the model reference point);\textit{IfcDirection} attribute of \textit{IfcSite} (stores rotations regarding project or global north).\tablefootnote{Ad-hoc solution used by several tools.} \\ \hline
		\textit{LoGeoRef40} & Any Cartesian CRS, including projected coordinates (CRS not specified in the file) & Attribute \textit{WorldCoordinateSystem} storing the coordinates of the reference point in any Cartesian CRS (including the projected ones) and direction \textit{TrueNorth}. Both are stored within \textit{IfcGeometricRepresentationContext}.\tablefootnote{Most official IFC2x3-way to store the reference system.} \\ \hline
		\textit{LoGeoRef50} & Specific projected CRS, specified by means of the EPSG code & IFC v.4 only. Coordinates of the reference point stored in \textit{IfcMapConversion} using the attributes \textit{Eastings},	\textit{Northings} and \textit{OrthogonalHeight} for global elevation. Rotation for the XY-plane stored using the attributes \textit{XAxisAbscissa} and \textit{XAxisOrdinate}. The CRS used is specified by \textit{IfcProjectedCRS} in the attribute \textit{Name} by means of the proper EPSG code.  \\ \hline
	\end{tabular}
	\caption{List of georeferencing options in IFC classified as \textit{LoGeoRefs} \citep{clemen2019level}. EPSG is a coding list for CRS.\tablefootnote{\url{https://epsg.io/}}}%
	\label{tab:LoGeoRefs}
\end{table}

The georeferencing of BIM has not been a priority for designers and software developers. Therefore the georeferencing information is not regularly stored and used, and not always read and exported in a completely consistent way. Moreover, the available tools to georeference BIMs are not perfectly optimized to acquire and store the correct information in the foreseen attributes and entities in the IFC files. For this reason, the IFC georeferencing, which is fundamental for GeoBIM integration, was investigated within the benchmark, from two main points of view.

On the one hand, the ability of software to correctly interpret the georeferencing information provided within the datasets was checked, with the focus on the georeferencing level LoGeoRef20 and LoGeoRef30. Since LoGeoRef30  is more accurate, allowing the storage of Cartesian projected coordinates, we asked participants to check if this one could be read and interpreted by software correctly, within the Task 1 (see Section~\ref{sec:metbench}).

On the other hand, it is quite unexplored if and how the tools can edit the georeferencing information of IFC models:
normally there is no or only little control of how georeferencing information is stored in the final model. For this reason, we tested this issue separately, within the benchmark Task 2 (see Section~\ref{sec:metbench}, \citep{noardo2020tools}).

\section{Methodology}

\subsection{The GeoBIM benchmark general set-up}\label{sec:metbench}

The benchmark was intended as a way to combine the expertise of many people with different skills, coming from several fields and interests, in order to describe the present ability of current software tools to use (\ie\ read, visualize, import, manage, analyse, export) CityGML and IFC models and to understand their performance while doing so, both in terms of information management functionalities, and possible information loss.
Moreover, since the large size of such standardised datasets often generates difficulties in their computational management, the ability to handle large datasets was a further part of the tests.

In particular, the four topics investigated in the benchmark are:

\begin{description}
	\item[Task 1]   What is the support for IFC within BIM (and other) software?\footnote{This Task is the object of this paper.}
	\item[Task 2]	What options for geo-referencing BIM data are available?
	\item[Task 3]	What is the support for CityGML within GIS (and other) tools?
	\item[Task 4]	What options for conversion (software and procedural) (both IFC to CityGML and CityGML to IFC) are available?
\end{description}

For this purpose, a set of representative IFC and CityGML datasets were provided~\citep{Noardo19a} and used by external, voluntary, participants in the software they would like to test in order to check the support in it for the considered open standard~\citep{Noardo3DGeoInfo2019a}.

Full details about the tested software and a full list of participants can be found in the respective pages of the benchmark website\footnote{\url{https://3d.bk.tudelft.nl/projects/geobim-benchmark/software.html} for the tested software and \url{https://3d.bk.tudelft.nl/projects/geobim-benchmark/participants.html} for the list of participants.}. The significant number of participants, balance in skills, fields of work, levels of confidence about the tested software (asked them to be declared) offered the possibility to limit the bias in the results.

The participants described the behaviour of the tested tools following detailed instructions and delivered the results in a common template with specific questions, provided as online forms.
In the end, they delivered both their observations and the models as re-exported back to the original standardised format (CityGML or IFC).

In order to cover the widest part of the list of software potentially supporting the investigated standards, we completed the testing ourselves, by searching the online documentation of both the standards and the potential software.

In the final phase of the project, the team coordinating the study analysed the participants' observations, descriptions and delivered further documentation (screenshots, log files, related documents and web pages).
From this review, an assessment of the performances and functionalities of the tested tools was derived.
Moreover, the delivered models were validated and analysed using available tools, when possible, and/or through manual inspection (Section~\ref{sec:mettask13}).
This approach allowed the inquiry about the level of interoperability given by the standard and its software implementation, by comparing the results of the export with the imported model features.

It is important to notice that the test results are not intended to substitute the official documentation of each software.
Moreover, there were no expertise nor skill requirements to participate in the benchmark tests. Therefore, some information could be wrong or inaccurate, due to little experience with the tested software or the managed topics.
The declaration of the level of expertise was intended to lower this possible bias.
Moreover, the benchmark team and the authors tried to double check the answers (at least the most unexpected ones) as much as possible, but the answers reported in the data were generally not changed from the original ones.
The eventual discrepancies between the best potential software performances and what was tested could anyway be showing a low level of user-friendliness of tools (and thus a degree of difficulty in achieving the correct result).

\subsection{The provided IFC datasets}

We have selected IFC datasets allowing the testing of the main issues regarding the support of IFC (\Reftab{tab:IFCdata}):
one dataset was the architectural IFC model of a small two-floor building, intended to test the support for IFC features in the most reliable way (\textit{Myran.ifc}). The second one was a large and complex architectural IFC model of a big tower in Rotterdam (\textit{UpTown.ifc}), intended to test the software's performance;
and the last one was the architectural IFC model of a medium-size building in IFC4 (\textit{Savigliano.ifc}), which was chosen to test also the support for IFC4 files and eventual discrepancies with the management of the other datasets, which were provided in IFC version 2x3 (which our previous experience suggests is better supported). All the IFC files where originally modelled in \textit{Autodesk Revit} and thereafter exported as IFC files.

Moreover, since performing tests of IFC geometries could be tricky in complex files combining many geometry types in an uncontrolled manner, we chose to monitor this aspect better by means of synthetically generated geometries (using \textit{IfcOpenShell}\footnote{\url{http://ifcopenshell.org}}) to test the support of software for specific kinds of geometry modelling (more detail in \Refsec{sec:IFCgeometries}). Those geometries are provided in both IFC2x3 and IFC4.

\begin{table}[H]
\small
	\centering
	\begin{tabular}{|p{1.7cm}|p{4.3cm}|p{1cm}|p{1.5cm}|p{2.2cm}|p{4cm}|}
		\hline
		\textbf{Name} & \textbf{Description} & \textbf{IFC vers.} & \textbf{Size (MB)} & \textbf{Source} & \textbf{Aim} \\ \hline
		\textit{Myran.ifc} & Model of a small 2-floor building in Sweden, by Swedish architects. Georeferenced. & 2x3 & 27.14 & Mondo arkitekter, (SE)
		& Test of the main functionalities of software and common procedures. \\ \hline
		\textit{UpTown.ifc} & Model of a big complex tower in Rotterdam, by Dutch architects. & 2x3 & 241.04 & Municipality of Rotterdam (NL) & Test of the software's performance. \\ \hline
		\textit{Sa\-vi\-glia\-no.ifc} & Model of a building in Italy, by an Italian architect within a research environment. & 4 & 21.55 & Arch. Lorenzo Polia (IT) 
		& Test of the support for IFC v.4 and to enable the tests of procedures and tools working with IFC v.4. \\ \hline
		\textit{Specific IFC geometries} & Set of geometries modelled using a range of the modelling alternatives allowed in IFC (see section \ref{sec:IFCgeometries}). & 2x3; 4 & 0.31 (v.2x3); 0.27 (v.4)  & Synthetically generated with \textit{IfcOpenShell}.
		& Test of the support and behaviour of software with respect to these specific geometries. \\ \hline
	\end{tabular}
	\caption{Summary of descriptions, features and aims of the provided IFC datasets.}%
\label{tab:IFCdata}
\end{table}

\subsubsection{IFC geometry sets}\label{sec:IFCgeometries}

\begin{figure}[H]
	\centering
    \includegraphics[width=0.8\linewidth]{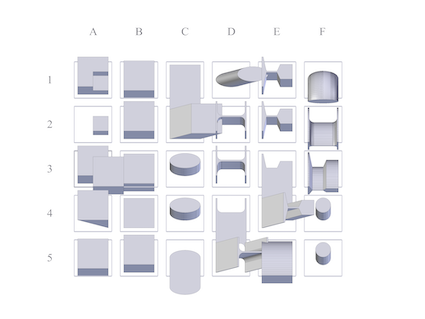}
    \caption{The dataset with synthetically generated IFC geometries.}%
    \label{fig:ifcgeometries}
\end{figure}

The geometries used in the BIM models can have a huge numbers of variations and a complete validation of them, as used in models, is still an unsolved task.
Moreover, IFC admits a number of geometry types that can be useful to modellers, but are sometimes not supported or can be interpreted in different ways by software.
On the other hand, IFC puts validity constraints on certain geometries, but some software implements workarounds aiming at reading those invalid geometries too, which are often undocumented, and there is therefore often little possibility to keep track of these solutions.
For these reasons and in order to explore the issues detailed in Section~\ref{sec:introIFCgeom}, a specific set of geometries  was provided among the benchmark datasets (\Reffig{fig:ifcgeometries},\Reftab{tab:ifcgeometries}).

\LTcapwidth=\textwidth
{
\newcommand{\VALID}{}
\newcommand{\INVAL}[1]{\textsuperscript{*#1}}
\newcommand{\YESINIFCFOUR}{}
\newcommand{\NOTINIFCFOUR}{\textsuperscript{\textdagger}}
\newcommand{\DEPTHNORMALIZED}{}
\newcommand{\DEPTHNONNORMALIZED}{\textsuperscript{\textdaggerdbl}}
\newcommand{\YESINCV}{}
\newcommand{\NOTINCV}{\textsuperscript{\textsection}}
\newcommand{\HASFILLET}{\textsuperscript{\textbardbl}}
\newcommand{\NOFILLETS}{}

\footnotesize
	\begin{longtable}{p{1.0cm}p{4.5cm}p{11cm}}
		
		\toprule
		\textbf{}   & \textbf{IFC Definition}      & \textbf{Description}                                                            \\ 
        \midrule
		\endfirsthead
		\endhead
		\textbf{A1}\NOFILLETS{}\VALID{ }\YESINCV{}\YESINIFCFOUR{} & IfcBooleanResult\_1          & Result of boolean subtraction with two cube operands with partial overlap                           \\ 
		\textbf{A2}\NOFILLETS{}\VALID{ }\YESINCV{}\YESINIFCFOUR{} & IfcBooleanResult\_2          & Result of boolean intersection with two cube operands with partial overlap                          \\ 
		\textbf{A3}\NOFILLETS{}\VALID{ }\YESINCV{}\YESINIFCFOUR{} & IfcBooleanResult\_3          & Result of boolean union with two cube operands with partial overlap                                 \\ 
		\textbf{A4}\NOFILLETS{}\VALID{ }\YESINCV{}\YESINIFCFOUR{} & IfcBooleanClipping\-Result\_1  & Result of boolean clipping operation with a cube and a halfspace operand                          \\ 
		\textbf{A5}\NOFILLETS{}\VALID{ }\YESINCV{}\YESINIFCFOUR{} & IfcShellBasedSurface\-Model\_1 & A shell based surface model, an explicit collection of faces                                      \\ 
		\textbf{B1}\NOFILLETS{}\VALID{ }\YESINCV{}\YESINIFCFOUR{} & IfcFacetedBrep\_1            & A faceted boundary representation, an explicit collection of faces                                  \\ 
		\textbf{B2}\NOFILLETS{}\VALID{ }\YESINCV{}\YESINIFCFOUR{} & IfcExtrudedAreaSolid\_1      & Extrusion of a rectangular profile                                                                  \\ 
		\textbf{B3}\NOFILLETS{}\INVAL{1}\YESINCV{}\YESINIFCFOUR{} & IfcExtrudedAreaSolid\_2      & Extrusion of a rectangular profile, negative extrusion depth                                        \\ 
		\textbf{B4}\NOFILLETS{}\INVAL{1}\YESINCV{}\YESINIFCFOUR{} & IfcExtrudedAreaSolid\_3      & Extrusion of a rectangular profile, zero extrusion depth                                            \\ 
		\textbf{B5}\NOFILLETS{}\VALID{ }\YESINCV{}\YESINIFCFOUR{} & IfcExtrudedAreaSolid\_4      & Extrusion of a rectangular profile, non-normalized direction vector                                 \\ 
		\textbf{C1}\NOFILLETS{}\INVAL{2}\YESINCV{}\YESINIFCFOUR{} & IfcExtrudedAreaSolid\_7      & Extrusion of a rectangular profile, direction vector parallel to profile                            \\ 
		\textbf{C2}\NOFILLETS{}\VALID{ }\YESINCV{}\YESINIFCFOUR{} & IfcExtrudedAreaSolid\_10     & Extrusion of a rectangular profile, slanted direction vector                                        \\ 
		\textbf{C3}\NOFILLETS{}\VALID{ }\YESINCV{}\YESINIFCFOUR{} & IfcExtrudedAreaSolid\_13     & Extrusion of a elliptical profile                                                                   \\ 
		\textbf{C4}\NOFILLETS{}\VALID{ }\YESINCV{}\YESINIFCFOUR{} & IfcExtrudedAreaSolid\_16     & Extrusion of a elliptical profile, non-normalized direction vector                                  \\ 
		\textbf{C5}\NOFILLETS{}\INVAL{2}\YESINCV{}\YESINIFCFOUR{} & IfcExtrudedAreaSolid\_19     & Extrusion of a elliptical profile, direction vector parallel to profile                             \\ 
		\textbf{D1}\NOFILLETS{}\VALID{ }\YESINCV{}\YESINIFCFOUR{} & IfcExtrudedAreaSolid\_22     & Extrusion of a elliptical profile, slanted direction vector                                         \\ 
		\textbf{D2}\HASFILLET{}\VALID{ }\YESINCV{}\YESINIFCFOUR{} & IfcExtrudedAreaSolid\_25     & Extrusion of a I-shape profile                                                                      \\ 
		\textbf{D3}\HASFILLET{}\VALID{ }\YESINCV{}\YESINIFCFOUR{} & IfcExtrudedAreaSolid\_28     & Extrusion of a I-shape profile, non-normalized direction vector                                     \\ 
		\textbf{D4}\HASFILLET{}\INVAL{2}\YESINCV{}\YESINIFCFOUR{} & IfcExtrudedAreaSolid\_31     & Extrusion of a I-shape profile, direction vector parallel to profile                                \\ 
		\textbf{D5}\HASFILLET{}\VALID{ }\YESINCV{}\YESINIFCFOUR{} & IfcExtrudedAreaSolid\_34     & Extrusion of a I-shape profile, slanted direction vector                                            \\ 
		\textbf{E1}\NOFILLETS{}\VALID{ }\NOTINCV{}\NOTINIFCFOUR{} & IfcExtrudedAreaSolid\_37     & Extrusion of a crane rail (A-shape) profile                                                         \\ 
		\textbf{E2}\NOFILLETS{}\VALID{ }\NOTINCV{}\NOTINIFCFOUR{} & IfcExtrudedAreaSolid\_40     & Extrusion of a crane rail (A-shape) profile, non-normalized direction vector                        \\ 
		\textbf{E3}\NOFILLETS{}\INVAL{2}\NOTINCV{}\NOTINIFCFOUR{} & IfcExtrudedAreaSolid\_43     & Extrusion of a crane rail (A-shape) profile, direction vector parallel to profile                   \\ 
		\textbf{E4}\NOFILLETS{}\VALID{ }\NOTINCV{}\NOTINIFCFOUR{} & IfcExtrudedAreaSolid\_46     & Extrusion of a crane rail (A-shape) profile, slanted direction vector                               \\ 
		\textbf{E5}\NOFILLETS{}\VALID{ }\YESINCV{}\YESINIFCFOUR{} & IfcRevolvedAreaSolid\_1      & Revolution of a rectangular profile                                                                 \\ 
		\textbf{F1}\NOFILLETS{}\VALID{ }\YESINCV{}\YESINIFCFOUR{} & IfcRevolvedAreaSolid\_2      & Revolution of a elliptical profile                                                                  \\ 
		\textbf{F2}\HASFILLET{}\VALID{ }\YESINCV{}\YESINIFCFOUR{} & IfcRevolvedAreaSolid\_3      & Revolution of a I-shape profile                                                                     \\ 
		\textbf{F3}\NOFILLETS{}\VALID{ }\NOTINCV{}\NOTINIFCFOUR{} & IfcRevolvedAreaSolid\_4      & Revolution of a crane rail (A-shape) profile                                                        \\ 
		\textbf{F4}\NOFILLETS{}\VALID{ }\NOTINCV{}\NOTINIFCFOUR{} & IfcSweptDiskSolid\_1         & Swept disk                                                                                          \\ 
		\textbf{F5}\NOFILLETS{}\INVAL{3}\NOTINCV{}\NOTINIFCFOUR{} & IfcSweptDiskSolid\_2         & Swept disk with parameter range outside of curve definition                                         \\ 
        
        \bottomrule
		
		\caption{Description of each object included in the IFC geometries set, in both IFC 2x3 and IFC 4 (the leftmost column refers to the row and column in \Reffig{fig:ifcgeometries}). \\
        \INVAL{1} not valid due to violation of the \textbf{positive} length measure requirement on \textsc{IfcExtrudedAreaSolid.Depth} \\
        \INVAL{2} not valid due to violation of the where rule \textsc{ValidExtrusionDirection} \\
        \INVAL{3} not valid as the parameter range for the sweep is outside the parametric range of the curve \\
        \NOTINIFCFOUR{} not included in the IFC 4 dataset due to removed entities in the schema. \\
        \DEPTHNONNORMALIZED{} valid, but using a non-normalized extrusion depth which may render differently in applications \\
        \HASFILLET{} this shape should have fillet radii, but not all viewing software displays those 
        }%
        \label{tab:ifcgeometries}
\end{longtable}}

The structure of the results template guiding the test about such datasets is as follows, for every geometry in the set.
First there is a general question whether an object is displayed in that slot after import.
Then what follows are questions about the positioning relative to the \(Z=0\) plane (mostly to distinguish how the negative extrusion depth is handled), the rendering of curved surfaces and what shape is displayed (to identify some specific situations such as the non-normalized extrusion direction).

\subsection{Answers analysis about the support for IFC}\label{sec:mettask13}

The methodology for analysing the results about the support of software for IFC (Task 1) and CityGML (Task 3) are very similar, since they were also conceived to test similar issues concerning interoperability and the ability of software to keep files consistent with themselves after import-export phases.

The initial part of results analysis (Section~\ref{sec:swsupptask1}) is qualitative, providing the description of software support and functionality based on the delivered answers.

The complete answers and documents delivered in the online templates\footnote{\url{https://doi.org/10.5281/zenodo.3964445}} \citep{noardofrancesca2020-3964445} 
were double checked for correctness and consistency with respect to the asked questions.
However, due to the nature of the initiative, we trusted the delivered information about the software, double checking it with new tests only in cases of inconsistent answers in different tests about the same software, or possibly, unexpected answers.
In these cases, we also considered the level of expertise of the participant to assess if further checks were actually needed.

The delivered answers in the templates were critically assessed, cross-checking them with the different tests about the same software and the attached screenshots. A score about each aspect considered for the assessment of general support and software functionalities is assigned, as: 1-full support; 0.5-partial support; 0- no support. Those are synthesized (\Reftab{fig:task1synthesis}), in order to more easily deduce possible patterns across many issues for a single software package or across many software packages for a single issue.

The definition of software groups are getting increasingly fuzzy, since the functionalities of all of them are continuously being extended and now tend to overlap with each other.
However, in the tables, and more generally, in the analysis, in order to help the detection of possible patterns, the tested software are classified considering the criteria that usually guide the choices made by users, based on their different needs for specific tasks:

\begin{itemize}
	\item \textit{GIS} are expected to combine different kinds of geodata and layers and make analysis on them, structured in a database, in a holistic system;
	\item \textit{`Extended' 3D viewers} are likely software that were originally developed for visualising the 3D semantic models, including georeferencing, and query them. They were (sometimes later) extended with new functions for applying symbology or making simple analysis.
	\item \textit{Extract Transform and Load (ETL)} software, and conversion software, are expected to apply some defined transformations or computations to data;
	\item \textit{3D modelling tools} have good support for geometry editing, but is not originally intended to manage georeferenced data nor semantics;
	\item \textit{Analysis software} are intended specifically for few kinds of very specific analysis (\eg\ energy analysis);
	\item \textit{BIM} software, are intended to design buildings or infrastructures according to the BIM method.
\end{itemize}

The investigated issues, reflected in the different sections of the provided templates, regarded mainly the support of the software for the two standards (how the software read and visualise the datasets) and the functionalities allowed by the software with standardised datasets (what is it possible to do with such data).
In particular, the test about the support was intended to check: how is the georeferencing information in the files read and managed (Section~\ref{sec:resloadgeoref}); how are the semantics read, interpreted and kept after the import (Section~\ref{sec:loadsemantics}); and how is the geometry after the import (Section~\ref{sec:loadgeom}). 

The last questions were related to the kind of functionalities (Section~\ref{sec:swfunctask1}) that are offered by the software for the management of IFC data:

\begin{itemize}	
	\item What kind of visualization is enabled (3D, 2D, with textures, with specific themes);
	\item What kind of editing is possible (attributes, geometry, georeferencing);
	\item What kind of query (query the single object to read the attributes, selection by conditions on attributes, spatial query, computation of new attributes);
	\item What analysis are allowed. This topic is more complex, since very different analysis can be possible. Therefore we summarized it by defining two analysis types: `Type 1' is any kind of analysis regarding the model itself (like geometric or semantic validation), and `Type 2' are the simulations and analysis about the performances of the represented object (\eg\ a building) with respect to external factors, in the city or environment (\eg\ shadow, noise, energy, etc.).
	\item Final issue: Is it possible to export back to IFC\@?
\end{itemize}

Moreover, the support for each of the delivered datasets were noted, given the specific features: IFC2x3 building model (Myran.ifc), IFC4 building model (Savigliano.ifc), very heavy model (UpTown.ifc).

This first parts provide a reference about the tools themselves for people intending to use standardised information.
In addition, the most challenging tasks and most frequent issues for the management of standards were supposed to be pointed out.

A second, more quantitative, part of the analysis considers the delivered models exported back to IFC from the tested software (Sections~\ref{sec:interoptask1Myran} and~\ref{sec:interoptask1Savigliano}).
The exported models of \textit{UpTown.ifc} were visually inspected but could not be completely analysed since the used tools (such as the \textit{NIST IFC analyser}) crashed without producing any useful result, probably given the big size of these datasets. For this reason the results regarding \textit{UpTown.ifc} are not described in the paper. The models can be though downloaded together with the other data from \citet{noardofrancesca2020-3964368}\footnote{\url{https://doi.org/10.5281/zenodo.3964368}}.

The numbers and types of features of such files were calculated and compared to the same features in the initial datasets that were provided for the test.
The \textit{NIST IFC analyser}\footnote{\url{https://www.nist.gov/services-resources/software/ifc-file-analyzer}}, calculating the number of IFC entities of each type in the dataset, was used for this check.
Moreover, manual inspection of the files within 3D viewers (\textit{BIM Vision, RDF IfcViewer}) was used to check the apparent problems (\eg\ missing elements) and further changes (\eg\ grouping in storeys, transformation of building elements in \textit{IfcBuildingElementProxy}, change to some kind of element to others and so on).
The inspection of the text format was also helpful to check some elements (\eg\ the more formal ones, documenting organization, applications, IFC version and so on).

This allowed us to assess the level of interoperability that the connected standards-tools can actually reach in the different cases: \ie\ can the data be imported and re-exported without any change?

A further assessment (Section~\ref{sec:swperftask1}) was intended to evaluate the software and hardware connected performance.
The times declared by the testers were compared for each datasets to see how much their computational weight could affect their management within software.

Given the complexity of measuring software performance to the closest second, this was not requested of the users. Instead, they were asked to provide an approximate timing value for each test, according to a classification that was proposed following the way they could affect the perception or the work of a user, as explained in the following list:

\begin{itemize}
 \item It is almost immediate (good!)
 \item Less than a minute (ok, I will wait)
 \item 1--5 minutes (I can wait, if it is not urgent)
 \item 5--20 minutes (in the meantime I do other things)
 \item 20 minutes--1 hour (I cannot rely on it for frequent tasks)
 \item more than 1 hour (I launch my process and go home, definitely ineffective for regular work)
\end{itemize}

Other options included reporting if the software crashed or if the task was not possible with the software provided, and participants were also asked to provide information about the specification of the machine, as this may impact overall performance of the software.

\section{Results: support of software for Industry Foundation Classes}

\subsection{Tested software against support for Industry Foundation Classes}\label{sec:swtask1}

For the benchmark we tested 31 different software packages (\Reftab{tab:swtask1}), trying to cover all the possible solutions to manage IFC, although a huge number of software packages and other tools exist, considering that the BIM-related industry is quite developed and still growing. 
In the table, they are organised based on the kind of software and divided into: open source, proprietary and freeware (but not open source) software.
Moreover, the levels of expertise of participants making the tests (from Level 1 the least to Level 4 the most expert) are also reported.

Some pieces of software were tested several times, with different levels of expertise: 2 tests with \textit{eveBIM} (Level 1, beginner and Level 4, developer); 2 tests with \textit{FZK viewer} (Levels 1 and 2); 3 tests with \textit{Safe Software FME} (all experts); 2 tests with \textit{FreeCAD} (Levels 1 and 2); 3 tests with \textit{ArchiCAD}, Levels 1 and 2; and many tests about \textit{Revit}, as expected, since very popular software (Levels 1, 2 and 3).

In addition, more (unsuccessful) tools were considered, for example within the \textit{Tekla} suite, \textit{Tekla Structure} is supposed to be the only tool supporting IFC (and finding the trial versions of others was anyway tricky); \textit{Autodesk Fusion}, which is however not supposed to support IFC\@; \textit{bimspot}\footnote{\url{https://bimspot.io}} crashed when trying to import any of the datasets.

Some other software, found on the internet or initially proposed by the participants that were found to be unable to support IFC were \textit{ACCA Solarius} and \textit{Solarius PV}, \textit{iTown}, \textit{BimView} and \textit{bimvie}.
There were also issues with finding the trial versions of other software, such as \textit{BuildingReconstruction}, \textit{OpenDesignAlliance IFC SDK}, only allowing some of the tools to be tried, or not reading IFC correctly (\eg\ other \textit{Bentley} software, \textit{ACCA EdiLus}).
However, the most popular tools were covered and tested thoroughly.

\begin{table}[H]
\small
	\begin{tabular}{|p{3cm}|p{3cm}|p{7cm}|p{3cm}|}
		\hline
		& \textbf{Open Source} & \textbf{Proprietary} & \textbf{Freeware} \\ \hline
		\textbf{GIS Software} & & Bentley Map Enterprise [L1]

		ESRI ArcGIS Pro [L3] & \\ \hline

		\textbf{`Extended' 3D viewers} & Datacomp Sp. z.o.o. BIM Vision [L1]

		RDF IFC Viewer [L1] & CSTB eveBIM [L4]
		
		CSTB eveBIM Viewer [L1]

		TeamSystem STR Vision IFC Viewer [L1]

		Solibri Anywhere [L1] & FZKViewer [L1+L2]  \\ \hline
		
		\textbf{ETL and conversion software} & & FME Desktop [3L3] & \\ \hline
		\textbf{3D modeling software (CAD)}  & FreeCAD [L1+L2]

		Bricsys Blender [L1] & Bentley MicroStation+TerraSolid [L3]

		Trimble SketchUp [L1]

		Cadwork Lexocad [L1]

		Bricsys BricsCAD Ultimate [L1] & \\ \hline
		
		\textbf{Analysis software} & & ACCA PriMus-IFC [L1]
		
		Dlubal RFEM [L1] & \\ \hline
		\textbf{BIM software} & BimServer [L2] & Autodesk Civil 3D [L1]

		Solibri Office [L1]

		Tekla Structures [L1]
		
		ACCA usBIM.viewer+ [L1]
		
		CadLine Ltd ARCHline.XP [L1]

		Simplebim [L1]

		Allplan [L1]

		AutoCAD Architecture [L1]

		ACCA Edificius [L1]

		Autodesk Revit [3L3+3L1+1L2]

		Vectorworks Designer [2L1]

		Graphisoft ArchiCAD [2L1+L2]

		DDS-CAD [L1]

		Autodesk Infraworks [L1] & \\ \hline

	\end{tabular}
\caption{Software tools tested against support for IFC within the benchmark Task 1. The participant Level 1 of expertise is abbreviated L1, etc.}%
\label{tab:swtask1}
\end{table}

\subsection{Software support for IFC}\label{sec:swsupptask1}

In \Reftab{fig:task1synthesis}, the more qualitative analysis of the participants' reports\footnote{\url{https://doi.org/10.5281/zenodo.3964445}} \citep{noardofrancesca2020-3964445} 
is synthesized. The result is discussed below focusing on the themes georeferencing, semantics and geometry.

\begin{sidewaystable}[htbp]
	\centering
	\includegraphics[width=1\linewidth]{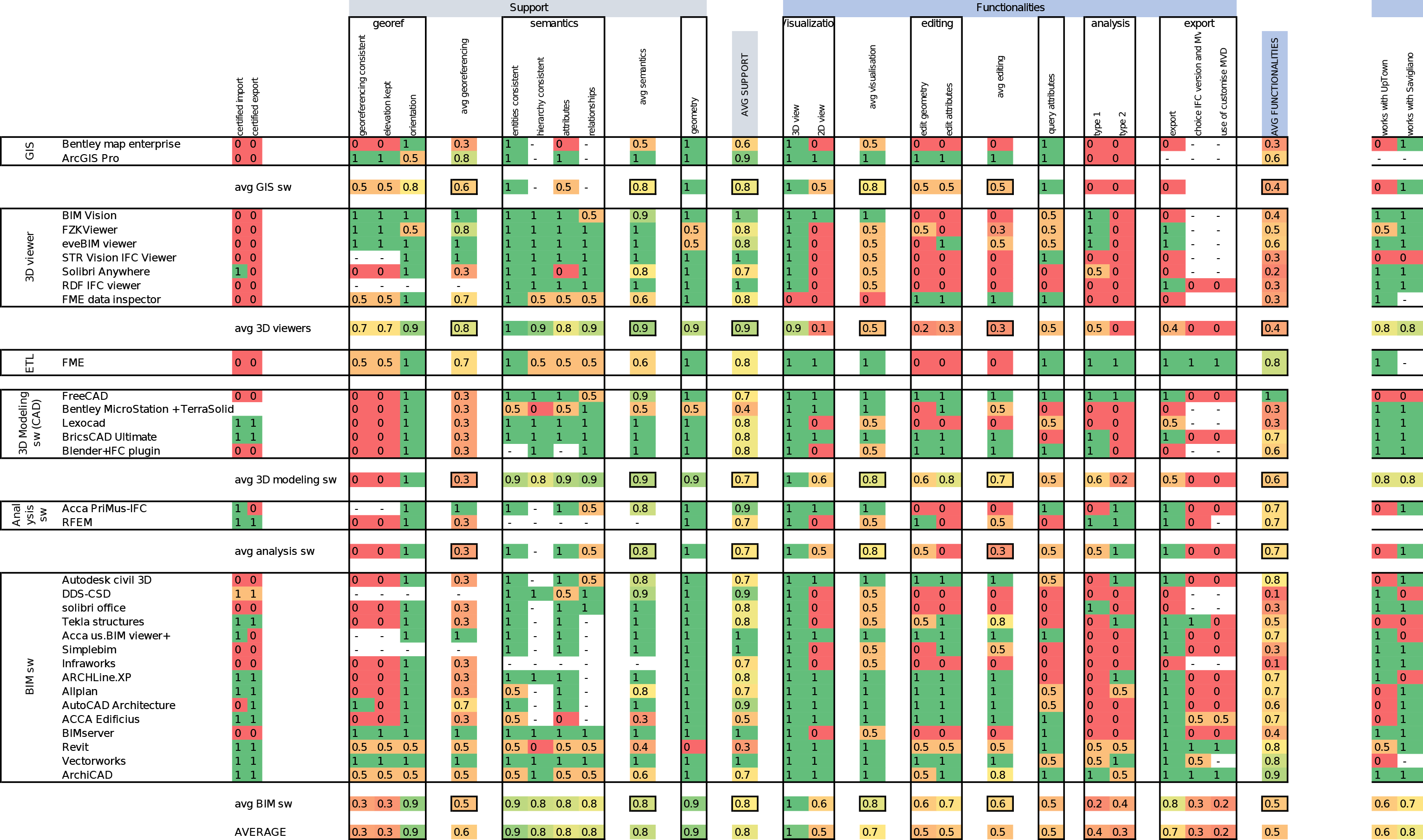}
	\caption{Synthesis table of the delivered tests regarding support for IFC (benchmark Task 1). The colour scale from green to red is assigned according to the scores from 1-full support to 0-no support.}%
	\label{fig:task1synthesis}
\end{sidewaystable}

\subsubsection{Load of IFC data: Georeferencing information}\label{sec:resloadgeoref}

It is possible to notice that in most of cases, georeferencing information, as well as elevation, is not read correctly and the models are moved to a local system having the origin in (0,0,0), using different units of measure (mainly metres and millimetres). In some cases it is possible to change this. For example in \textit{eveBIM}, by default, it would load the model with a local coordinate system; it is however possible to change the settings to have it use the model's previously defined local coordinate system.

Some tools work partially, for example, \textit{FME} uses \textit{RefLatitude} \textit{RefLongitude} in \textit{IfcSite}, although more accurate coordinates are assigned through an \textit{IfcLocalPlacement} (LoGeoRef30) or the project's representation context (LoGeoRef40).

In other cases, the CRS is misinterpreted, for example \textit{Bentley Microstation+Terrasolid}
moved all the models to 0,0,0 in the CRS EPSG:3152. In other software, some georeferencing information is reported in very high detail, including datum, ellipsoid and so on. However, this information is not correct.

With some other software, like \textit{Revit} or \textit{ArchiCAD} different tests give different answers, therefore it is possible that it depends on import settings.

To summarize, only 30\% of the tools correctly use the georeferencing information with proper global coordinates (as provided by LoGeoRef30 or 40).
Very few changes are reported about the tests with the IFC4 dataset, which  does not use the extended IFC4 georeferencing features anyway.
Orientation of the model is instead kept correctly in most of the cases (90\%), although it was difficult to assess if the added True North information was read.
 
\subsubsection{Load of IFC data: Focus on Semantics}\label{sec:loadsemantics}
 
The semantics are read more consistently than the georeferencing information, with general good support, although some inaccuracies are anyway found, mostly consisting in:
partial recognition of entities\footnote{As examples, in \textit{Bentley Microstation+TerraSolid}, for the \textit{Myran.ifc} model, only \textit{IfcWall} and \textit{IfcWallStandardCase} were translated; in \textit{ACCA Edificius} \textit{``opening, door, and curtain wall are detected by the software; IFCBeam and IFCCovering are loaded as an IFC Proxy Object; It also has vertical/horizontal envelopes which are actually not IFC classifications''}; for \textit{Revit}: \textit{``some classes that are different in IFC are the same in Revit, following the standard settings; it is possible to set own Revit Categories (therefore set them to the correct IFC Class name)''}; in \textit{ArchiCAD}, \textit{``Some classes/layers are missing, most notably are the roofs and windows''}; \textit{Tekla structures} does not import all the entities, although the present ones are consistent.};
loss (or partial loss) of relationships\footnote{In \textit{FME}, hierarchy and relationships are kept by means of parent ids; for \textit{Revit}, the loss of many relationships is reported, probably due to the need to separate elements while importing; same for \textit{BIM Vision}; in \textit{FreeCAD}, \textit{``spatial aggregation and containment is properly retained (although in debug mode, aggregations with more than 10 objects are skipped), Type objects (IfcRelDefinedByType) seem not to be considered, IfcRelConnectsPathElements (connection between walls) is ignored''}; In \textit{Autodesk civil 3D}, only storeys aggregation is kept; in \textit{ACCA PriMus-IFC}, it is only possible to find many \textit{``IfcRelConnectsPathElements''} relations.};
partial inconsistencies in the reading and interpretation or loss of attributes\footnote{In the \textit{FZK Viewer} sometimes attributes are different from the reference ones, although generally well-managed and interpreted; \textit{Bentley Microstation+TerraSolid} can read consistent attributes for the \textit{Uptown.ifc} dataset but cannot for \textit{Myran.ifc}; \textit{DDS-CAD} is supposed to read attributes correctly, however, the function did not work, therefore it was not possible to check; in \textit{Revit}, some of the attributes were consistent with the IFC ones, but most of them were missing; in \textit{ArchiCAD}, some objects are not recognized correctly in terms of attributes and semantics.}.

Approximately, 80\% of the tools manage semantics in a satisfactory way, especially the 3D viewers group performed well. However, many inaccuracies need to be solved to reach a good interoperability. In particular, it is noticeable how \textit{Revit} reports inconsistencies with respect to the imported original data, although being the software where all the datasets were modelled and from where they were exported.

\subsubsection{Load of IFC data: Focus on Geometry}\label{sec:loadgeom}

The software tools in this study are considered to support the geometry if there are no apparent errors; that is no detailed study of geometry was conducted on the building models, since this is a  very complex issue. However, reported errors generally regard the loss of some elements or their change so that intersections, deformations or modifications (\eg\ in normals) occurred.\footnote{Specific reported features include:
\textit{FZK Viewer}, generally works well, except with the \textit{Myran.ifc} model, whose geometrical representation looks a bit distorted in one of the tests, while the geometry is not visualized at all in the other one; Moreover, with the \textit{Savigliano.ifc} (IFC4) file, some elements (roof pitches) are not visualised;
\textit{eveBIM} detected wrong normals, although we do not know if they were wrong in the beginning or were changed during import, 
and with respect to IFC4 file, the \textit{IfcAdvancedBrep} is not implemented so the geometry is not shown in this case;
about \textit{Bentley MicroStation +TerraSolid}, some slight changes in dimensions are reported;
\textit{BIMServer} seems to read geometry correctly, except that the roof at the top of the tower is missing, and possibly windows;
about \textit{Revit} issues are reported: some walls intersecting with the floors and beams that do not join correctly, subtracting the volumes; mistakenly display of subtraction solids inside the families of doors and windows, which are used to pierce the walls; often the stratigraphy of walls is not correct in the corners and intersections;
in \textit{Vectorwork} a scale can be applied to the data, thus, it is necessary to check that it is correctly set.}

In addition, \Reftab{fig:task1synthesis} reports whether the software is able to load and work with the different datasets.
Issues include: failures most likely connected to the computational requirements in the case of \textit{Uptown.ifc}; to the IFC version in case of \textit{Savigliano.ifc}; and maybe to possible geometry issues in the case of \textit{Myran.ifc}, which is not read correctly in a few cases.

For example, the version 0.18 of \textit{FreeCAD} is not able to import the \textit{Myran.ifc} model, and one of the tests with \textit{FZK viewer} fails with \textit{Myran.ifc} as well. In addition, other import errors are reported in \Reftab{tab:task1errors}.

\begin{table}[H]
\small
	\begin{tabular}{|p{0.12\linewidth}|p{0.83\linewidth}|}
		\hline
		\textbf{Software} & \textbf{Import error} \\ \hline
		FZK Viewer & with Savigliano.ifc:\ The software crashes when trying to add an element

		with Uptown.ifc:\ Many errors reported

		with Myran.ifc:\ Error 776:\ Boolean Operation - clipping plane generation for CdgisModel geometry failed; Geometry face - Invalid outer loop; Geometry polyline - Polyline contains collinear points, points removed; Geometry polylines - Double point removed \\ \hline
		
		eveBIM & with Savigliano.ifc: some geometries can't be generated. Error message is: ``Warn - IFCProduct (\#565376): error IfcRepresentationItem (\#564617) Error - this representation Item IfcAdvancedBRep (\#564922) is not implemented'' (this will be a further development of the software) \\ \hline
		
		RFEM & Many elements were not imported because said to be ``not relevant for structural analysis''. \\ \hline
		
		Tekla structure & You have to convert it to Tekla's own structure first. It could not convert quite many objects. \\ \hline
		
		Allplan & It ignored 14 objects for Myran and 440 for Savigliano, but does not state which ones these are. \\ \hline
		
		AutoCAD Architecture  & Despite there not being any errors during the import, not all elements shown on the data page are present. \\ \hline
		
		ACCA Edificius & It reported the following kind of errors: incorrect element referencing, unexpected error while calculation Entity 2409 cross beam, a critical error has occurred in the structural model calculation. \\ \hline
		Revit & IFC opening elements not imported and needed to be discarded. \\ \hline
	\end{tabular}
\caption{Errors in the imports for some of the software tools.}%
\label{tab:task1errors}
\end{table}

\subsubsection{IFC geometries interpretation}\label{sec:geometriestask1interp}

\begin{figure}[H]
\centering
\includegraphics[width=\linewidth]{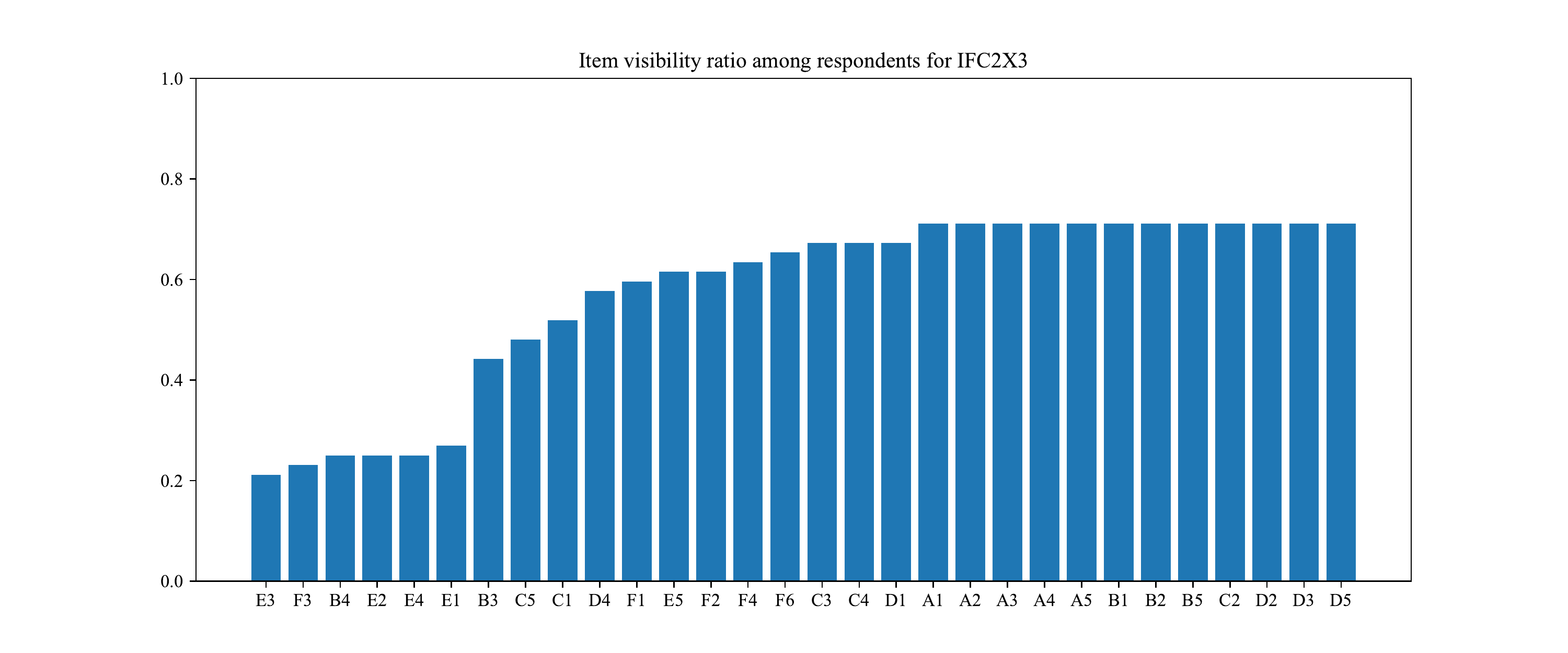}
\caption{Item visibility ratio for IFC2X3. The labels on the horizontal axes refers to the column and row of the geometry in \Reffig{fig:ifcgeometries}}%
\label{fig:ifc-ana-visibility-ifc2x3}
\end{figure}

\begin{figure}[H]
\centering
\includegraphics[width=\linewidth]{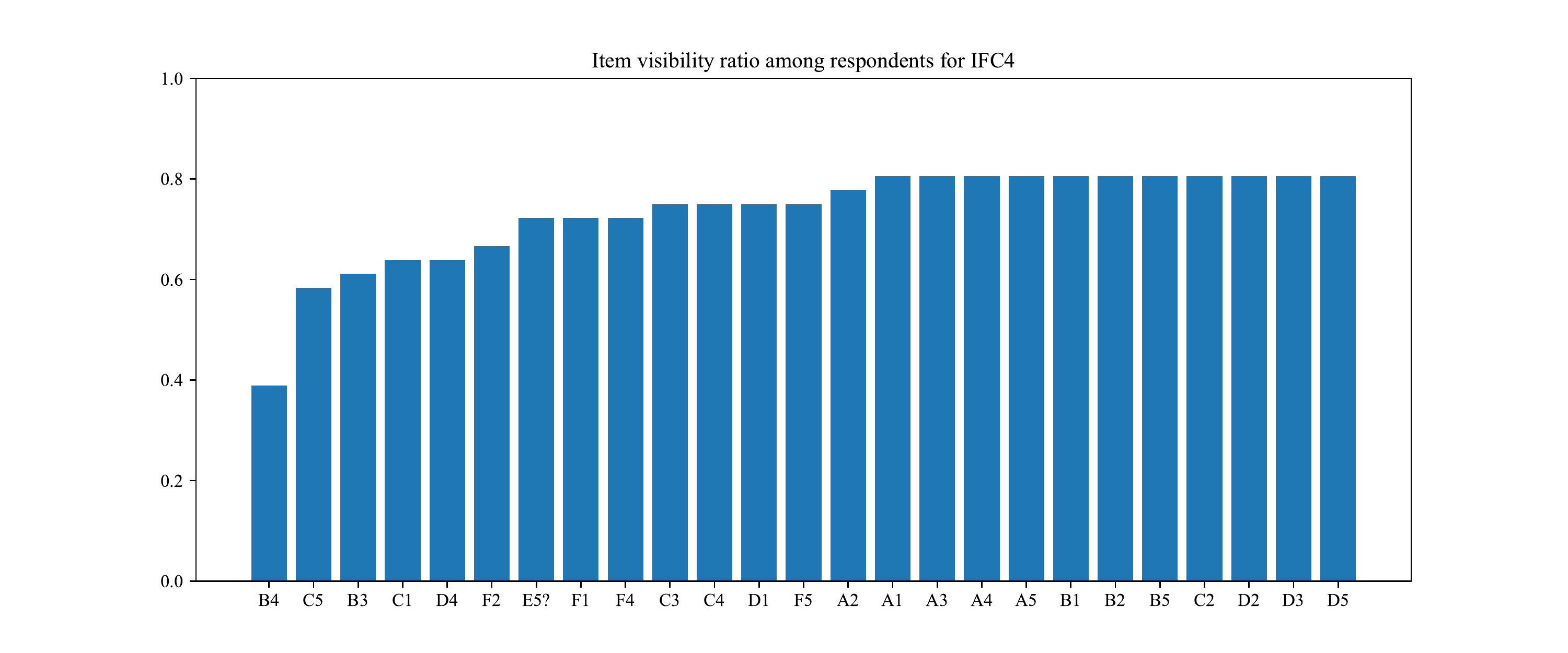}
\caption{Item visibility ratio for IFC4. The labels on the horizontal axes refers to the column and row of the geometry in \Reffig{fig:ifcgeometries}}%
\label{fig:ifc-ana-visibility-ifc4}
\end{figure}

The dataset is synthetically generated for both the IFC2X3 schema as well as the IFC4 schema, with the IFC4 geometries being a subset of the ones supplied for IFC2X3.
For that reason one would expect similar import success ratios for the IFC2X3 and IFC4.
This is not the case, as shown in Figures~\ref{fig:ifc-ana-visibility-ifc2x3} and~\ref{fig:ifc-ana-visibility-ifc4}.
A possible explanation is that applications supporting IFC4 are newer or better and therefore have higher success rates.

\begin{figure}[H]
	\centering
	\includegraphics[width=\linewidth]{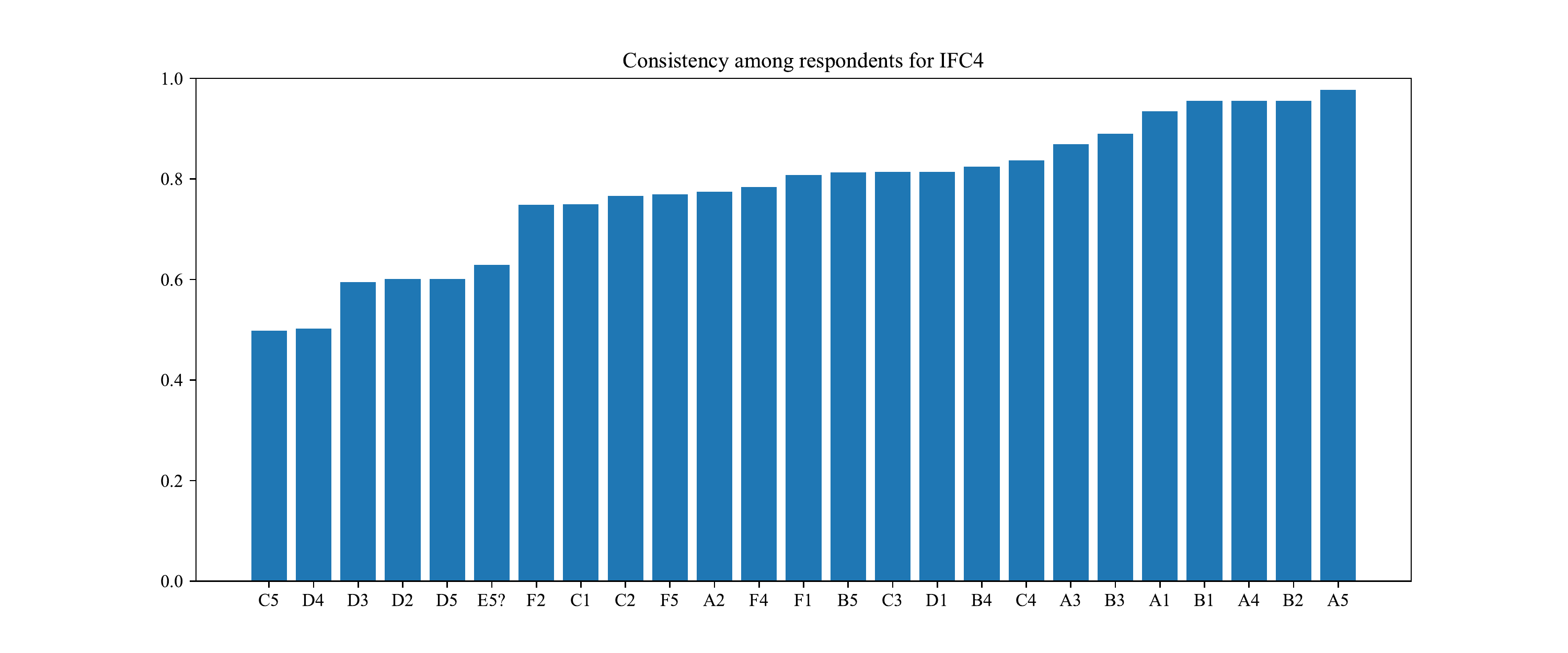}
	\caption{Consistency among respondents, defined as the pair-wise equality of all combinations of answers (1.0 in case all respondents answer the same, 0.0 when all different, although that's generally due to the nature of the questions). Consistency is averaged over the three follow-up questions (regarding position, surface shading and shape type) and excludes answers of the respondents that indicated the shape is not displayed.}%
	\label{fig:ifc-ana-consistency-ifc2x3}
\end{figure}

As mentioned in the discussion in earlier paragraphs, one of the main concerns of the authors regarding the state of IFC support in importing tools is the divergence in how situations are handled. On the one hand this can be explained as evolving support for more complicated geometry types or types that are not considered native to the domain of the importing tool. On the other hand, what can be observed in practice is that some tools import invalid situations without notifying the user.
In some cases these are due to tools not performing checks on the entity instance attribute values or on the resulting geometries.
For example, in the case of the negative extrusion depth, when the IFC parser does not validate WHERE rules automatically, detecting the negative extrusion depth requires an explicit statement in the program code that may have been omitted accidentally.
In other cases there is a deliberate effort to ``heal'' certain invalid situations.
For example the \textit{IfcOpenShell} software library provides functionality to detect self-intersections in face boundaries definitions and discard all but the largest cycle of edges\footnote{\url{https://github.com/IfcOpenShell/IfcOpenShell/blob/df81490/src/ifcgeom/IfcGeomFunctions.cpp\#L3656}}. While on the one hand this is implemented with good intentions, this behaviour does deviate from strict standard compliance and results in different flavours or dialects of IFC being supported among applications, which in the long run will hurt interoperability.
The divergence of importing applications is displayed in \Reffig{fig:ifc-ana-consistency-ifc2x3}.

The least consistency among the IFC2X3 results is seen in D4, C1, C5 (various profiles with invalid extrusion direction).
Naively one would expect consistent results here given that this is a violation of a schema constraint, but this shows the state of implementation in the industry that support for these constraints in parser libraries cannot be assumed, especially not the complicated ones as in this case (the \texttt{IfcDotProduct} function).
D3, D5, D2 are also inconsistent which stems from the question on whether fillet radii are incorporated and whether the surfaces are shaded smooth or faceted.
Interesting to highlight is that there is quite some consistency in F4 and F5 as virtually no application took the \textit{StartParam} and \textit{EndParam} of \textit{IfcSweptDiskSolid} into account.


\subsection{Using IFC data}\label{sec:swfunctask1}

BIM tools are usually very complex and making a very detailed report about them is difficult.
However, the general functionalities regarding visualization, editing, query, analysis and export possibilities are tested (\Reftab{fig:task1synthesis}).
All the software offer visualisation functions, mainly only 3D, usually also 2D within the BIM and 3D modelling software groups.
Those same two groups are the ones mainly allowing editing of geometry or attributes.
Usually, query functionalities are partial, within most of the tools\footnote{
They are foreseen within GIS software;
3D viewers usually have partial query abilities, for example \textit{BIM Vision} provides general pre-defined ways to view the information, listed by structure, by types, by groups (system, zones or other) and by layers. For each, properties, location, classification and relations can be shown at the object level; also \textit{FZK Viewer} offers mainly pre-defined queries;
\textit{Lexocad} can only query based on object type (\eg\ wall, slab, wall standard, \ldots);
in \textit{Autodesk civil 3D}, as well as in \textit{Allplan} and \textit{AutoCAD Architecture} it is only possible to select properties which are in the native software format;
in \textit{Vectorworks}, query tools are present, however query is possible if the project is connected to a database; object selection and information retrieval is possible;
finally, \textit{ArchiCAD} presents a ``find-and-select'' tool, but issues are reported.
}.

Looking at analysis, very few software packages allow it: only 20\% for the analysis concerning the model itself, Type 1, and 30\% analysing the designed building performances, Type 2, which is very little for something which is supposed to be an operational tool.
Since many different kinds of analysis can be within the two categories (Type 1 and 2), and the tested tools offered very various features, only the general support for the two types are reported in \Reftab{fig:task1synthesis}.
In some cases, the support is partial, for example, some Type 1 analysis are sometimes foreseen but not working, as it is the case of \textit{Solibri Anywhere}, or only in specific cases, like \textit{Vectorworks}, working only with NURBS\@.
With the Type 2 analysis the pattern is similar, with some tools giving issues, for example, in \textit{Allplan}, \textit{Revit} and \textit{ArchiCAD}, the Type 2 analysis sometimes do not work with the datasets, or with IFC files in general, although available.
It is quite difficult to give an exhaustive overview of this, though.

80\% of the software can export to IFC and only 30\% of them allow some customization about the IFC version and Model View Definition to be used.
Additional notes and warnings were reported to the export operations, regarding the possibility to customize the mapping of entities, in some cases, and a few errors about the way geometry and other features could result from the export.\footnote{
in \textit{FME}, there were a number of warnings related to inappropriate geometry types;
in \textit{FreeCAD}, the default settings yielded an error regarding colours (to be checked in later versions and reported for fixing if necessary). By changing the settings to use the IfcOpenShell serializer, the error did not appear;
for \textit{ACCA PriMus}, it is possible to add additional properties;
in \textit{Revit} it is possible to customize the export by means of manual entity mapping;
similarly in \textit{Vectorworks}, a layer mapping pre-process was needed to export the file. Each layer had to be selected separately and assigned to the correct ``vector story name''.
}

No sharp connection between the IFC certification and the software support and functionalities is verifiable.

\subsection{Writing of IFC2x3 files: analysis of the exported \textit{Myran.ifc} models}\label{sec:interoptask1Myran}



The models which were exported from the tested pieces of software \citep{noardofrancesca2020-3964368}\footnote{\url{https://doi.org/10.5281/zenodo.3964368}} 
were analysed by means of the \textit{NIST IFC analyser}, in order to formally check them.
The tool counts the number of entities, relationships and properties and summarizes it in reports.
From comparing the summaries, counting the features in the exported files, with the ones regarding the original datasets, we obtain general information about the way in which the software modified the models and we can point out which are the entities that tools have limited support for.

\begin{figure}[H]
	\centering
	\includegraphics[width=1\linewidth]{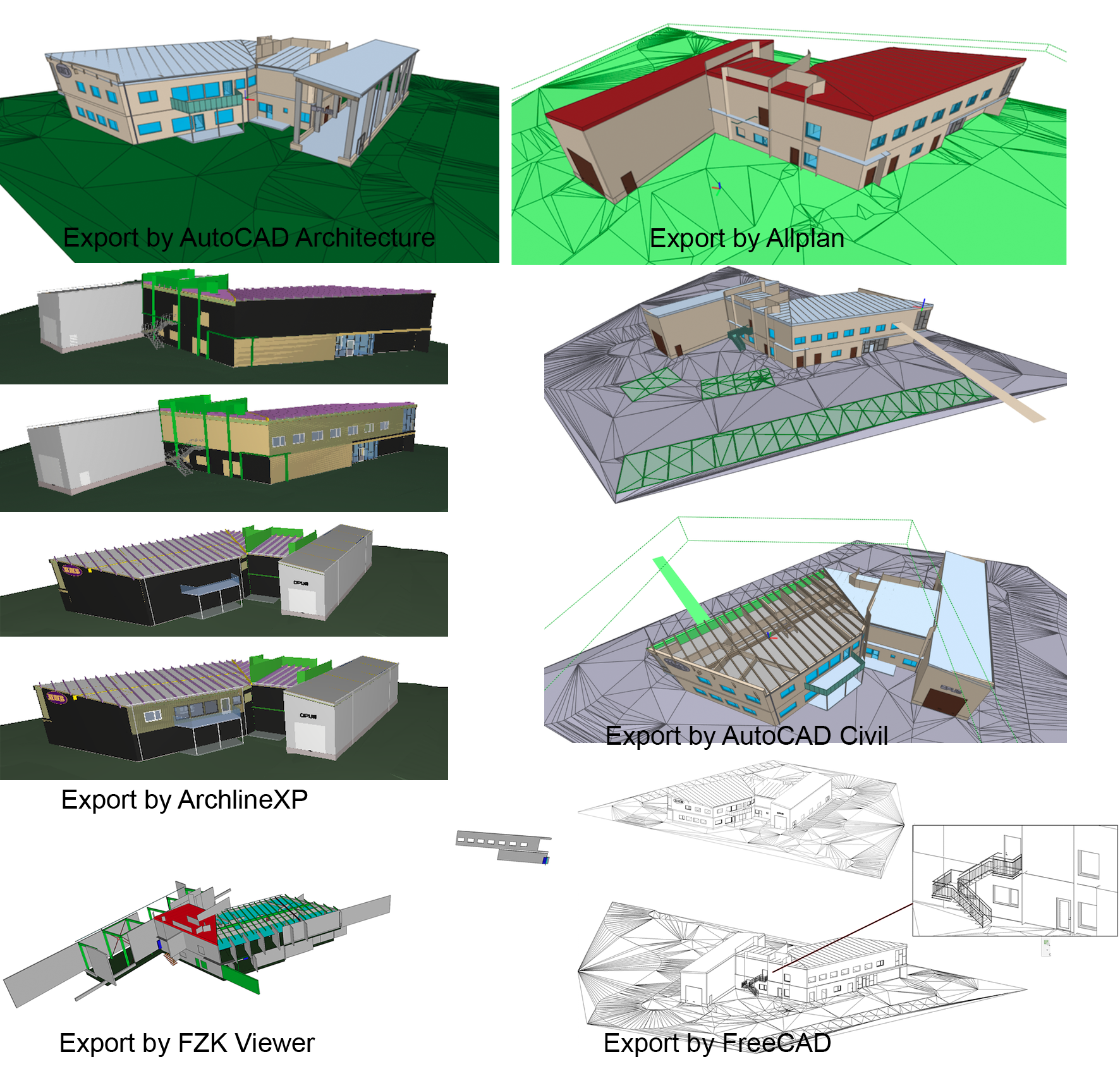}
	\caption{Exported Myran models.}%
	\label{fig:myranexport}
\end{figure}

Some of the delivered \textit{Myran.ifc} models were re-exported by the tools without errors, but apparently the export failed, since the building is not represented anymore: this is the case of one of the \textit{ArchiCAD} tests, where only the signboard of the building is stored; the test with \textit{Tekla structure}, where only one beam is there; and one of the models exported by \textit{Revit}, which is empty.
In addition, the models exported by \textit{FZKViewer} could not be read by \textit{BIM Vision} nor \textit{NIST IFC analyser} to be analysed and inspected, and \textit{RDF IfcViewer} only shows the \textit{IfcSite} geometry, as terrain, and other two random (wrong) objects in one of the \textit{FZKViewer}-exported cases, and a completely wrong building in the other one (\Reffig{fig:myranexport}).


In the models shown in \Reffig{fig:myranexport}, issues in the exported models mainly regard a change in the model properties, loss of elements, change in the grouping of entities and some occasional apparent change in geometry.\footnote{
The model exported by \textit{FreeCAD} gives some problems when trying to load it into viewers: \textit{RDF IfcViewer} can show it only very far and small (probably some change in the coordinate reference system happened), and it can be hardly handled in such software; \textit{BIM Vision} is not able to visualise it; \textit{Revit} gives a warning (``IFC:only 2 points in polyloop \#702627, expected >=3''), 
but then shows it quite correctly. 
There, it is possible to see that some elements are missing (\eg\ the beams in the exterior stairs).
In \textit{Allplan}, 
the difference that could be noticed with respect to the original is the lack of the external stair and, in the grouping, the IfcSite is stored separately than the three other storeys, which is actually a better storage, but anyway different from the original benchmark-provided file.
In \textit{AutoCAD Architecture,} the three storeys are collapsed in a single one, big parts of objects belonging to the side building are missing (\eg\ walls, doors), the roofs' colour changed for some reason, probably due to the loss of some properties, and some of the walls that were cut at the roof level, are not, so that they protrude above the roof. 
In \textit{AutoCAD Civil}, the three storeys are collapsed in only one, the terrain model constituting the \textit{IfcSite} is split in its different components (parking places, street, different parts of the terrain), which are recognizable by a human eye, but were not labelled differently in the original dataset. All of these are represented by means of \textit{IfcBuildingElementProxy} entities. Also in this case, the roofs and the walls are changed in the same way as in the \textit{AutoCAD Architecture} model. Moreover, one of the walls has a part protruding towards the exterior in a clear error.
In the model exported by \textit{Archline XP}, 
the windows and doors seem to be disappeared, even if they rest below the most exterior layers; probably the reason was some issues in the use of \textit{IfcOpenings} to subtract the volumes corresponding to doors and windows in all the concerned walls and coverings. The roof slab is also missing.}

The two \textit{Revit} ones, the two \textit{ArchiCAD} ones, the \textit{ACCA PriMus }one, the \textit{ACCA usBIM.viewer} one (only showing something weird, similar to a duplication in the wireframe representing the site), in \textit{eveBIM}, \textit{Simplebim}, \textit{FME} and \textit{Bricscad} look similar to the original dataset, and one could assume they are consistent.

However, a closer look into the kind and number of entities included in the models, by means of the NIST IFC analyser analysis (\Reftab{fig:MyranNIST}), shows they are actually not.

\begin{table}[htbp]
	\centering
	\includegraphics[width=0.9\linewidth]{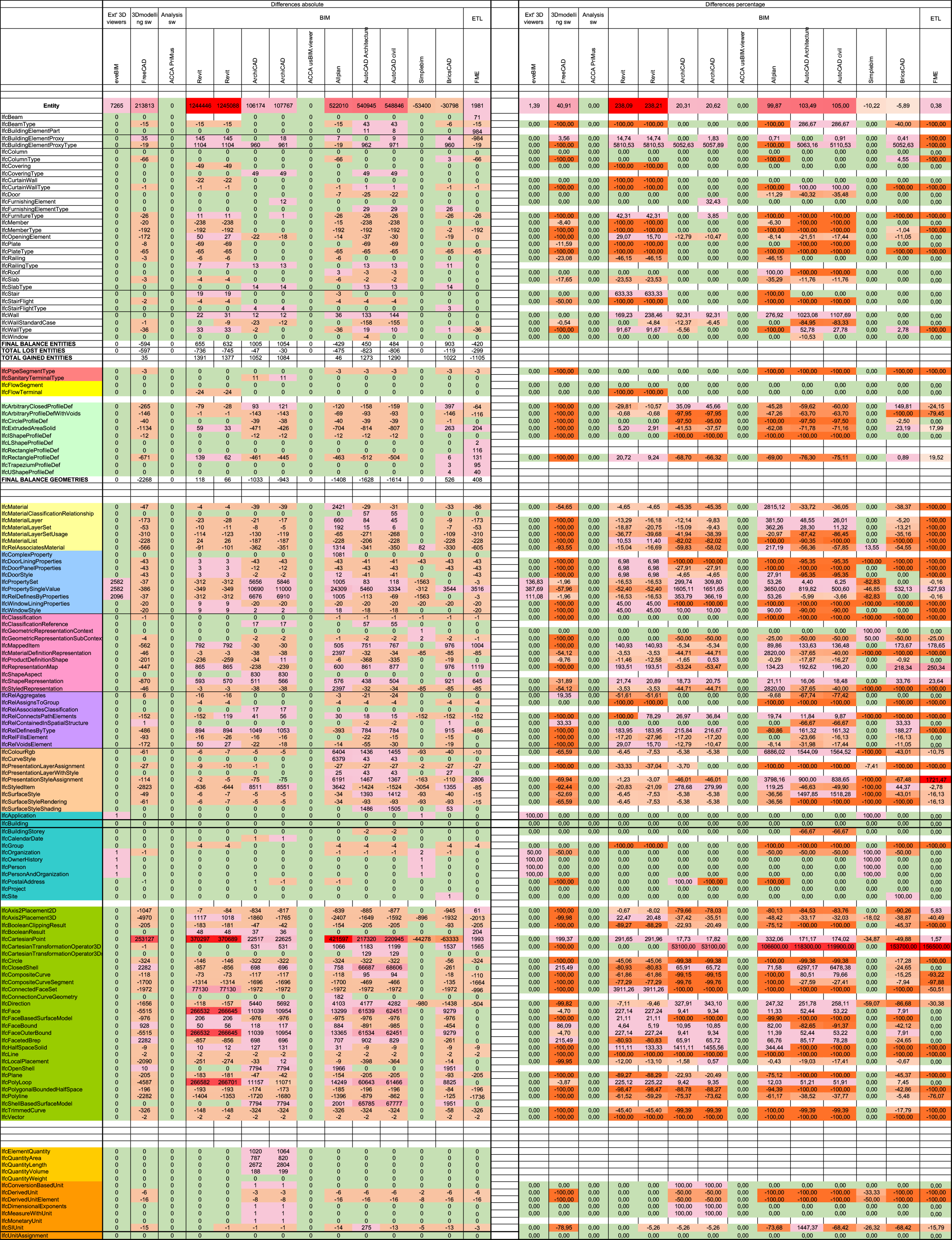}
	\caption{Differences in the NIST analysis results between the Myran models exported by the tools and the provided one.}%
	\label{fig:MyranNIST}
\end{table}

The only two tools leaving the model completely unchanged (same number of each feature) are: \textit{ACCA Primus}, which is a software supporting calculation of materials for costs assessment, and \textit{ACCA usBIM.viewer}. The only IFC entities which are left unchanged across all the models are \textit{IfcUnitAssignment} and \textit{IfcBuilding} (which is 1 for all the models).

The section regarding some general information, from \textit{IfcApplication} to \textit{IfcSite} (blue-green), is among the most consistent ones. It includes subcategories of \textit{IfcSpatialStructures} information, together with some meta-data level information. 
Apparently, it is quite easy for the software to keep track of such entities, interpreting and writing them back in a correct way: the number of \textit{IfcBuildings} remains the same (one).
Moreover, for example, \textit{eveBIM} and \textit{Simplebim} add one \textit{IfcApplication} to the file, probably intended to document that the file passed also through such tools.
Therefore it is not a mistake in the writing of the file, but a wise update of the file metadata.
The same applies for \textit{IfcOwnerHistory}, \textit{IfcOrganization}, \textit{IfcPerson}, \textit{IfcPersonAndOrganization}.
However, the syntax used to add the information in these two files is not the same.
Although both are not wrong, this flexibility in the writing of the new file would for sure be a difficult condition for the interpretation of the file by third applications.

In the section representing entities concerning quantities (dark yellow), 
which are not present in the original file, almost all the models remain the same, except for two of the \textit{ArchiCAD} tests (which report different numbers, though). 
Also the units representation (orange) is kept quite consistent, and the inconsistencies could be due to the different software settings for export (both for the original file and for the tested ones), not necessarily meaning inconsistency in the model.
\textit{IfcDerviedUnit} and \textit{IfcDerivedUnitElement} entities are the least kept ones by the re-exported models, which lose them completely in 7 out of 16 cases.

The most interesting part of this analysis is however concerning the IFC entities contained in the `Shared Building Elements' and `Product extension' parts of the IFC structure (white).

Models exported by the same software by different participants (\eg\ the two coming from \textit{Revit}, the two coming from \textit{ArchiCAD}) present very similar results in the number of entities, although not exactly the same even.
The same happens with \textit{AutoCAD Architecture} and \textit{AutoCAD civil}, which we can suppose using similar algorithms.

Moreover, it is not possible to find any balance in the lost/gained entities considering neither the groups of entities being subclasses and superclasses, or that can be considered as alternative representation of the same object (\eg\ \textit{IfcWall}, \textit{IfcWallStandardCase}, \textit{IfcWallType} or \textit{IfcStair}, \textit{IfcStairFlight}, \textit{IfcStairFlightType}), nor the specific entities and the \textit{IfcBuildingElementProxy} and \textit{IfcBuildingElementProxyType}.

It is also curious how some additional elements appear even if the similar ones are kept consistently.
For example, in the case of `stairs' entities, in both \textit{ArchiCAD} tests and in \textit{Bricscad}, although the \textit{IfcStairFlight} entities are still in the same number, other 4 \textit{IfcStairFlightType} objects (same number of occurrences of \textit{IfcStairFlights}) are added (only three in one of such cases, for some reason). 
In the models exported by \textit{Revit}, 19 more \textit{IfcStair} entities are added.

Similar reasoning is possible when considering \textit{IfcWall}, \textit{IfcWallStandardCase} and \textit{IfcWallType}, although in this case in (only) one of the \textit{ArchiCAD} models and in the \textit{Allplan} case the balance is kept, at least, with \textit{IfcWallStandardCases} becoming \textit{IfcWalls} in the \textit{ArchiCAD} case and \textit{IfcWallType} becoming \textit{IfcWalls} in \textit{Allplan}.
Such change in the choice of an entity whose meaning is similar, although losing the optimization advantages for which different entities were proposed in the data model, would keep the model semantically consistent.
However, it is not possible to verify a consistent behavior with respect to such choice in software.

The models exported by \textit{FreeCAD} lose parts of object (as we could see from the visual inspection, for example, part of the external stairs are no more represented), as well as \textit{Allplan} (\eg\ again the external stair).
Moreover, \textit{Revit}-exported files, \textit{FreeCAD}, \textit{Allplan}, \textit{AutoCAD Architecture}, \textit{AutoCAD Civil} lose all the \textit{IfcMemberType} entities.

But looking at the final balance in the number of entities, we can see that their number raises.
Therefore, assuming that no objects or parts of the model were not exported at all,
we can just report on a major number of entities describing the building elements, that means that probably some of the objects were split into several ones.
The increase in the amount of entities is also confirmed by the number of geometric objects, whose numbers vary substantially, losing many instances of several geometric entities, while some of them are more than doubled (especially the most simple ones, such as \textit{IfcCartesianPoint}, \textit{IfcFace} or \textit{IfcPolyLoop}).
Again, a pattern is difficult to be found.
Similarly, for the parts concerning the other groups of entities, it is difficult to find similarities across the software or the same entities.

The increase in the number of entities is also reflected in the file sizes, that change from the original 27 MB to a maximum 94 MB for both the \textit{Revit} models.

In addition, the size decreases to 24 MB in the models exported by \textit{eveBIM}, \textit{SimpleBIM}, \textit{Bricscad}.
This could mean that either those ones are better optimized than the original dataset or that something went lost in the conversions.
In both cases it does not depict the best situation for interoperability, even if it maybe not too bad per se.

\subsection{Writing of IFC4 files: analysis of the exported \textit{Savigliano.ifc} models}\label{sec:interoptask1Savigliano}

There are similar considerations for the exported \textit{Savigliano.ifc} models as for the \textit{Myran.ifc} model.

The models exported by \textit{BIMserver}, \textit{eveBIM} and \textit{SimpleBIM} look good.
Again, the models exported by \textit{FZKViewers} can't be opened in \textit{BIM Vision}, nor it is shown in the \textit{RDF IfcViewer} or analysed by the \textit{NIST IFC analyzer}.

\begin{figure}[H]
	\centering
	\includegraphics[width=1\linewidth]{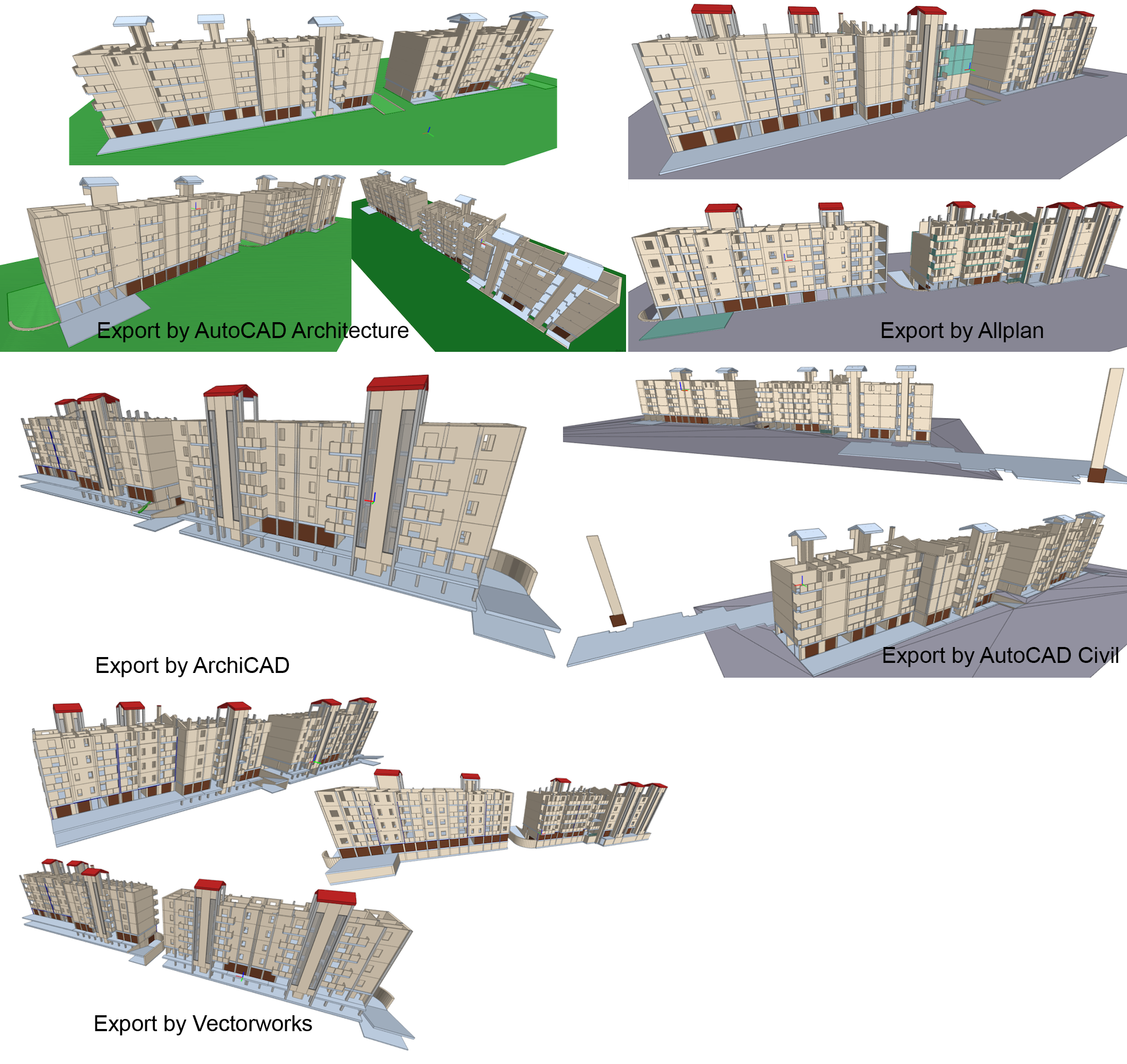}
	\caption{The exported Savigliano models.}%
	\label{fig:saviglianoexported}
\end{figure}

In the other models, the loss of elements and properties was the main issue, besides a few changes in some geometries\footnote{
In the model exported by \textit{Allplan}
, the upper part representing the roofs is missing, and all the windows as well.
Moreover some of the external walls and some slabs are coloured in a different way (blue), but it is not very clear what is changed.
The model exported by \textit{AutoCAD Architecture}
 looks very similar to this last one: the top part of the building is almost completely missing, even a bigger part than in the \textit{Allplan} model.
Moreover, the windows are missing and, in this case, the openings in the walls, where the windows and doors would be supposed to be inserted, are missing too, presenting plain filled walls.
Even the ramp leading to the underground parking is filled and appears as the continuation of the site.
Moreover, other elements are missing, such as the windows in the stair towers, closing elements.
In addition, the colour of some elements, especially slabs are missing, so that probably some property was lost.
%
%
The \textit{AutoCAD civil} one is very similar to the \textit{AutoCAD Architecture} one, but in addition, a slab and a door are completely displaced and the site colour changes. 
The ramp to access the underground parking is however well represented.
One possible explanation is that, being the software intended to represent civil works, those ramps can be better interpreted with respect to the same in \textit{AutoCAD Architecture}, although probably the two tools being based on similar algorithms.
%
%
The model exported by \textit{ArchiCAD} also misses the top part and windows.
Moreover, the site around the building is missing.
An apparent error is a sort of beam (classified however as wall) that goes from one of the ground slabs into the ramp going to the parking (in green in \Reffig{fig:saviglianoexported}).
%
%
Another model, exported by \textit{Vectorworks}, is very similar, although without the beam invading the ramp.}
(\Reffig{fig:saviglianoexported}).

When looking at the \textit{NIST} analysis
, it is again possible to notice several more inconsistencies, except for the model exported by \textit{BIMServer}, which leaves the model exactly unchanged.

\begin{table}[htbp]
	\centering
	\includegraphics[width=0.6\linewidth]{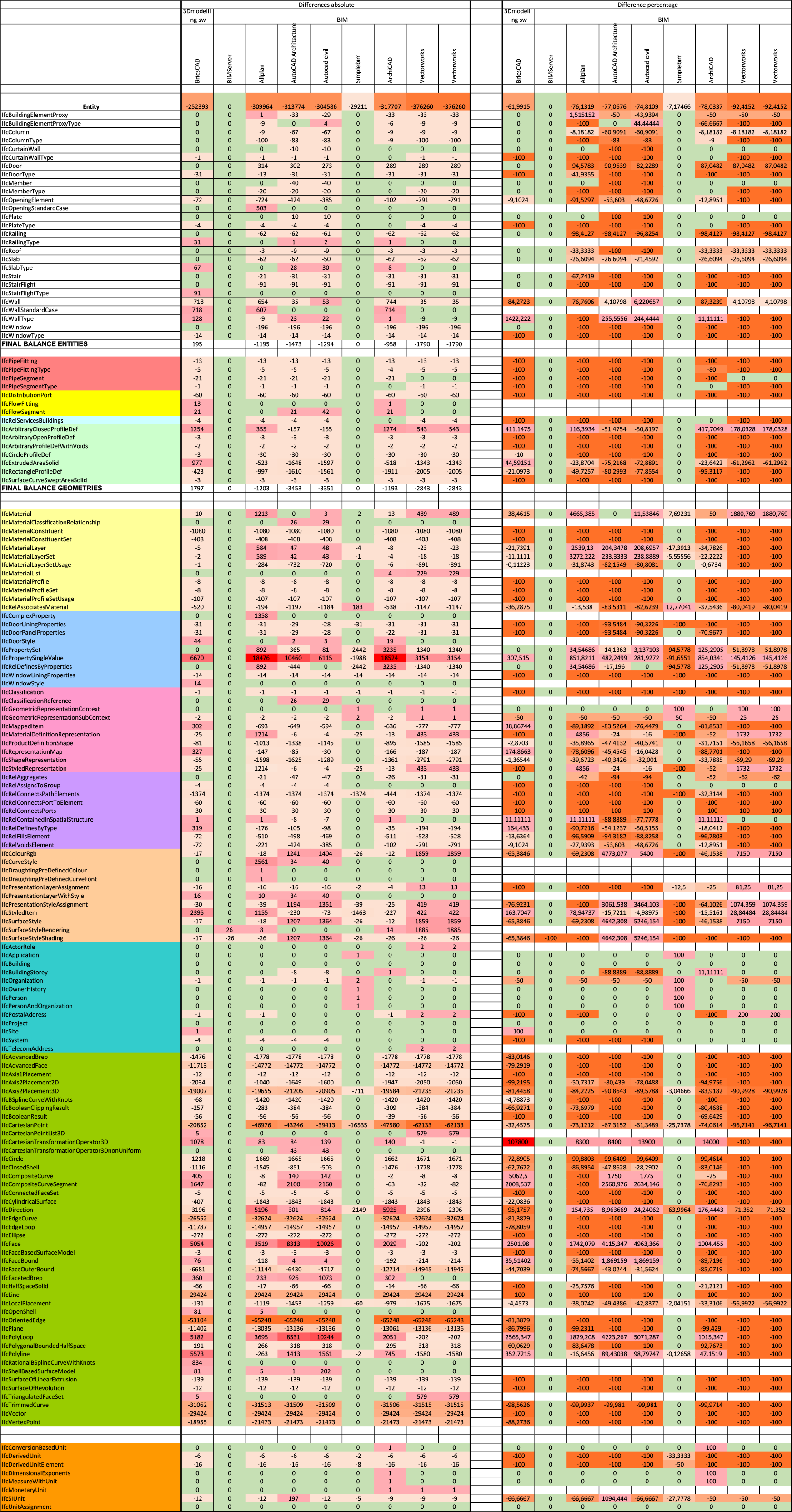}
	\caption{Differences in the NIST analysis results between the Savigliano.ifc models exported by the tools and the provided one.}%
	\label{fig:SaviglianoNIST}
\end{table}

\textit{Simplebim} comes second, considering that all the others perform very bad.
In this case there are no differences in the number of building elements, and very little difference among the geometric entities, where many \textit{IfcDirections}, 25\% of \textit{IfcCartesianPoints} and some other ones are lost.
In this case the main differences are found in properties, materials and relationships.

As in the case of \textit{Myran.ifc}, it is very difficult to find patterns in the loss/gain of entities.
In the model exported by \textit{Bricscad} it is possible to see 718 \textit{IfcWalls} lost and 718 \textit{IfcWallStandardCases} gained, which were probably resulting from the conversion of these ones.
The numbers do not correspond perfectly, but similar behaviour can be seen also in \textit{ArchiCAD} and \textit{Allplan}, while \textit{AutoCAD Architecture} loses 35 \textit{IfcWalls} and gain 23 \textit{IfcWallTypes}.
The most general entity being \textit{IfcWall}, among them, this is a weird result though, since the opposite direction of the conversion would be expected to be the safest one for the sake of consistency.

Very little other patterns can be found.
In this case many entities are lost, contrary to the \textit{Myran.ifc} case, maybe due to the little support for some IFC4 new entities.
We could also see from the visual inspection that many objects were lost and this is even clearer by the entities count, where we can see that many are lost completely.
The tendencies are generally not straightforward, even if it is possible to notice that the added entities are mainly \textit{IfcPropertySingleValue}, \textit{IfcFace}, \textit{IfcPolyLoop}, but the behaviour varies case to case, and it is necessary to look at the \Reftab{fig:SaviglianoNIST} to check the details.

\subsection{Writing of IFC geometries: analysis of the exported IFCgeometries.ifc and IFC4geometries.ifc models}\label{sec:interoptask1IFCgeom}

The differences in the NIST analysis results between the IFCgeometries.ifc models exported by the tools and the provided one is shown in \Reftab{fig:geomNIST} and those for IFC4geometries.ifc in \Reftab{fig:geomIFC4NIST}. The information in the two tables indicates that there is no relevant difference between the test with the dataset in IFC v.2x3 and the one in IFC v.4. In the case of these datasets, the original files consisted only of geometries, with only minimal semantic information. Therefore, when it is present in the exported files (e.g. Materials, Styles and property sets), it means that the software which exported them had attached such information based on its default settings.

\begin{sidewaystable}[htbp]
	\centering
	\includegraphics[width=1\linewidth]{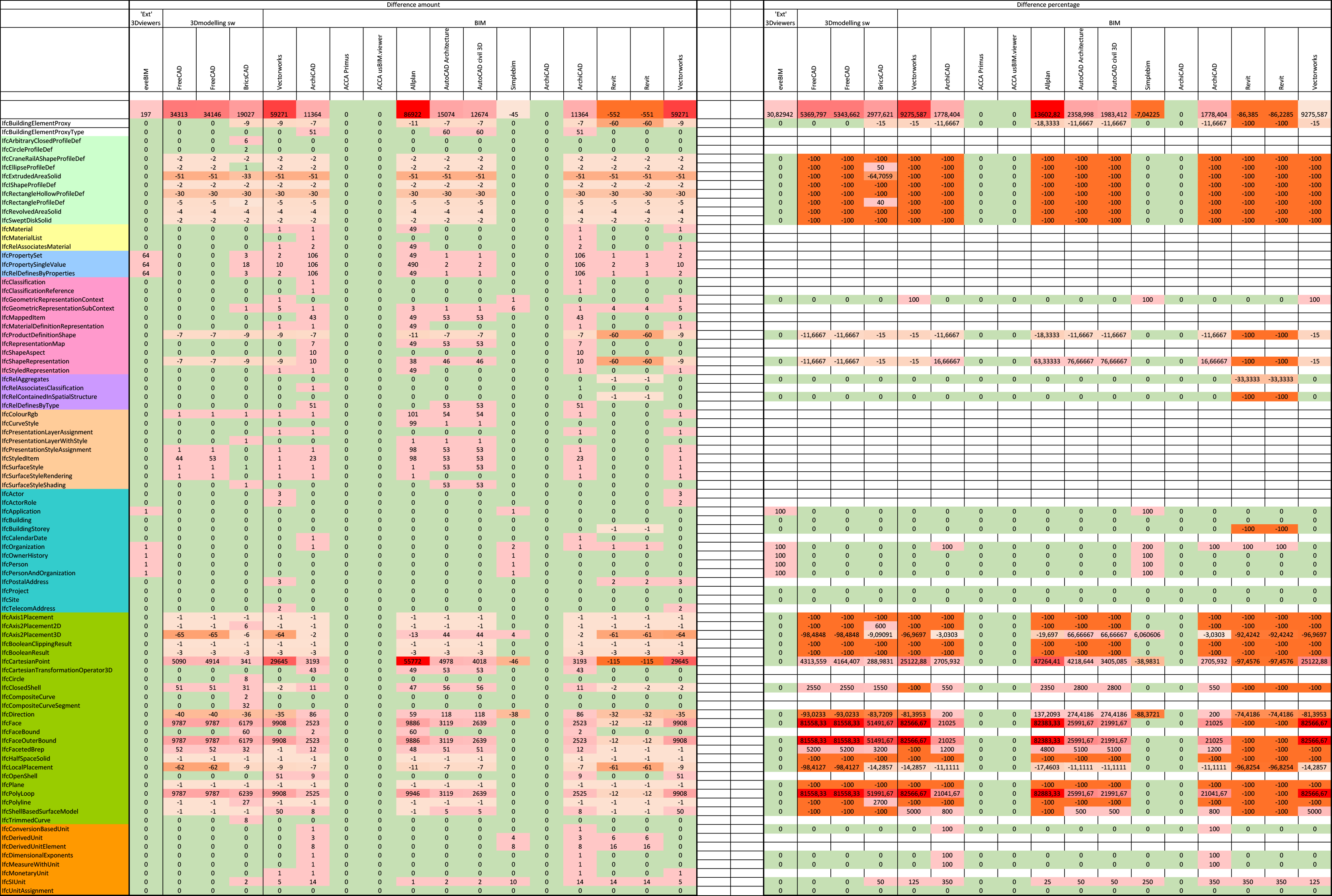}
	\caption{Differences in the NIST analysis results between the IFCgeometries.ifc models exported by the tools and the provided one.}%
	\label{fig:geomNIST}
\end{sidewaystable}

\begin{table}[htbp]
	\centering
	\includegraphics[width=1\linewidth]{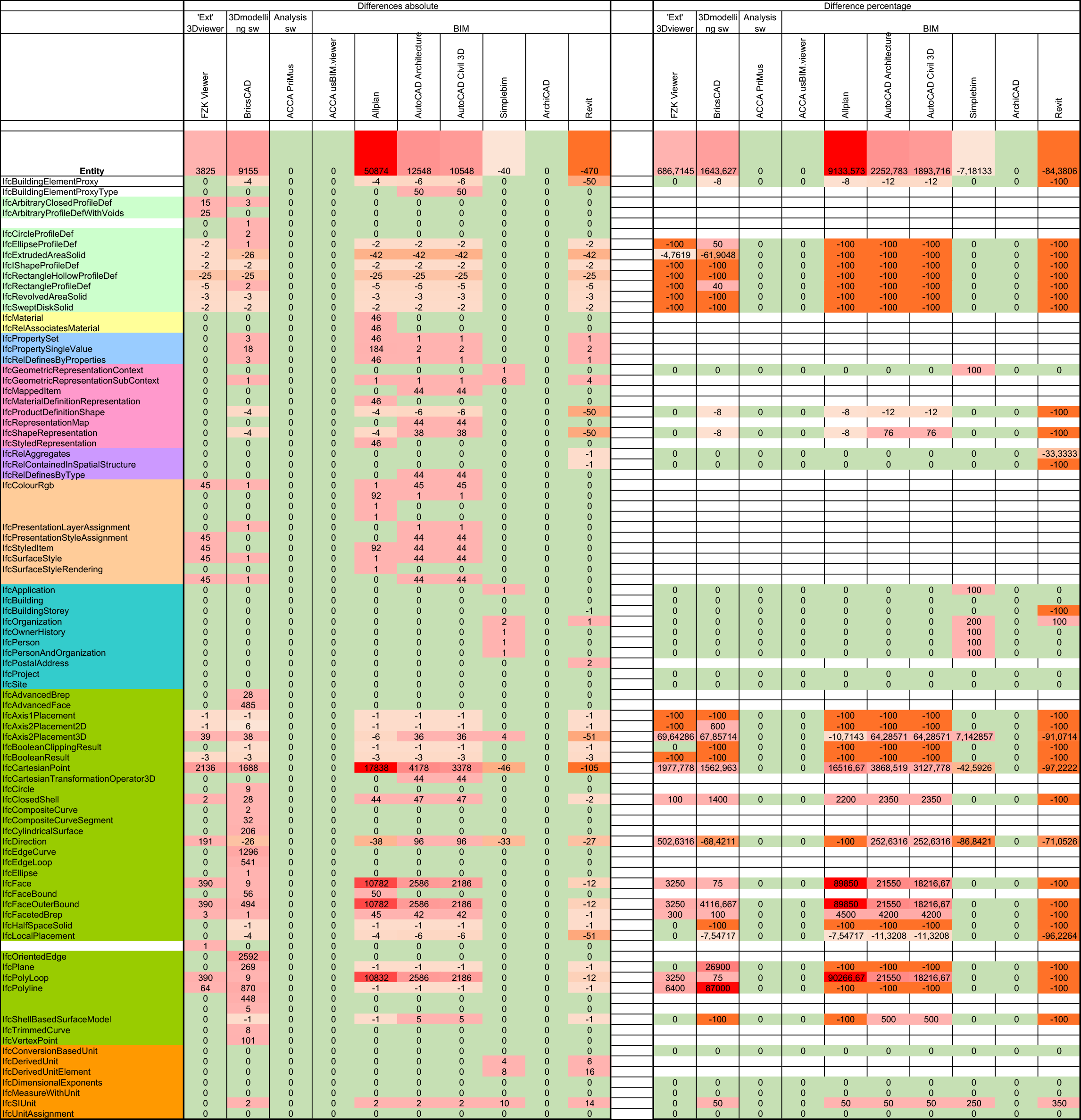}
	\caption{Differences in the NIST analysis results between the IFC4geometries.ifc models exported by the tools and the provided one.}%
	\label{fig:geomIFC4NIST}
\end{table}

Few tools are able to re-export the data leaving them completely consistent with the imported ones, namely: \textit{ACCA Primus},\textit{ ACCA usBIM.viewer} and \textit{ArchiCAD}. \textit{eveBIM} can also be included in the list of the good ones, since some elements are added regarding \textit{IfcPropertySets} and metadata, but no elements are lost. \textit{Simplebim} follows, with only few entities missing.

Many geometries resulting from parametric modelling are completely lost, as it is possible to see in correspondence of the lines in light green in \Reftab{fig:geomNIST} and \Reftab{fig:geomIFC4NIST} (e.g.\ \textit{IfcRevolvedAreaSolid}, \textit{IfcCircleProfileDef}, \textit{IfcIShapeProfileDef}, \textit{IfcSweptDiskSolid}) as well as some of the entities which are in the other green rows in the tables, such as \textit{IfcBooleanResults}.
Instead, it is possible to observe that the number of other more generic entities increases a lot, e.g.: \textit{IfcCartesianPoint}, \textit{IfcFace}, \textit{IfcFaceOuterBound}, \textit{IfcPolyLoop}, and so on.
Apparently, the parametric geometries tend to be simplified in the export, by most of the software, in favour of a more explicit representation.
It is also visible in the analysis of the other exported models, but in these datasets it appears very clearly.

\subsection{Software performances with IFC}\label{sec:swperftask1}

\begin{figure}[H]
	\centering
	\includegraphics[width=0.95\linewidth]{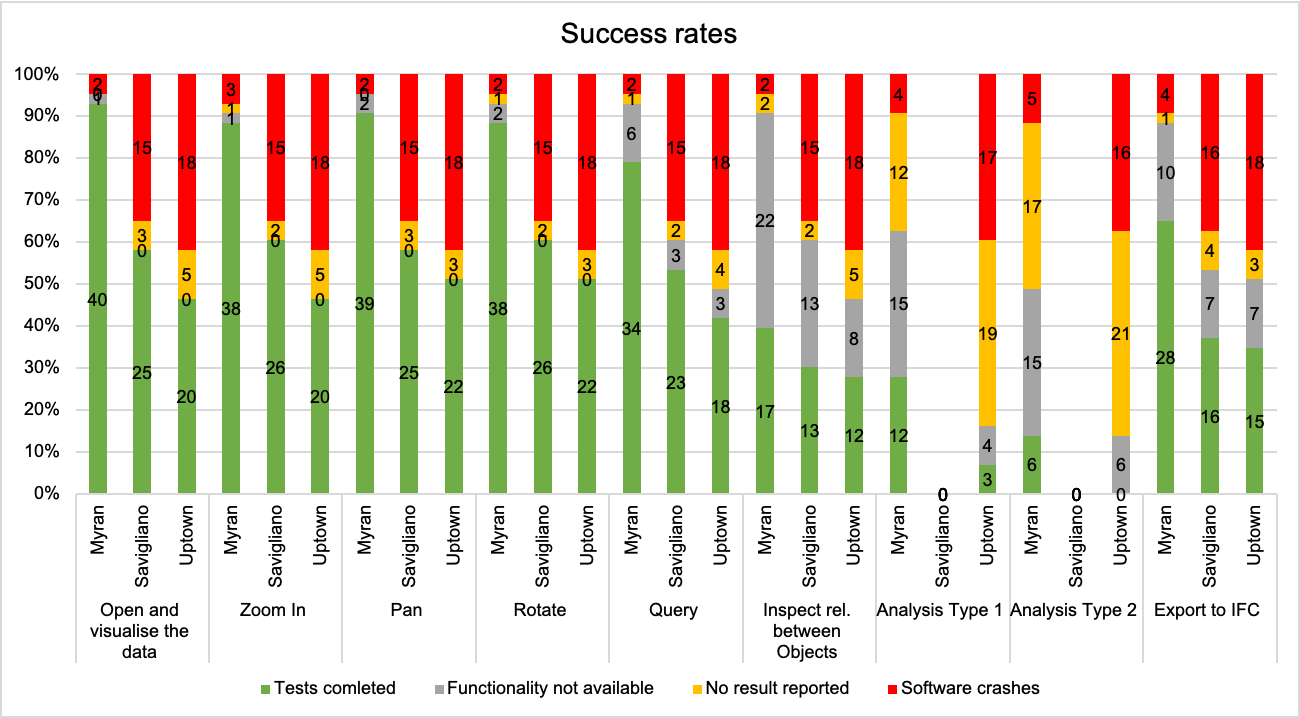}
	\caption{Graph reporting the number of tools associated to their success rate in the indicated tasks.}%
	\label{fig:successrates}
\end{figure}

\begin{figure}[H]
	\centering
	\includegraphics[width=0.95\linewidth]{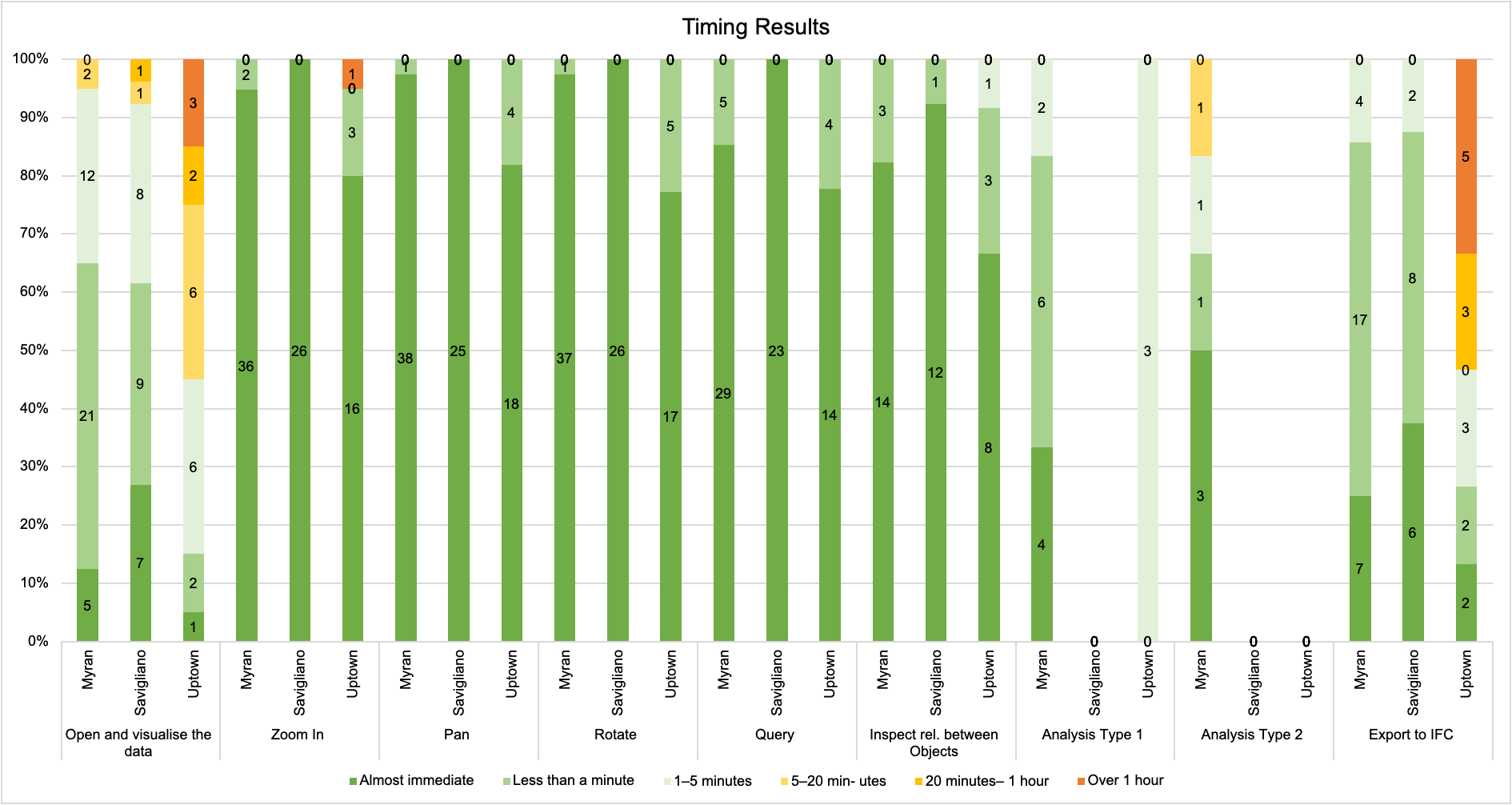}
	\caption{Graph reporting the number of tools associated to their timing results in the indicated tasks.}%
	\label{fig:timinggraph}
\end{figure}

A total of 43 different reports were returned, for 33 different software packages.
In particular, multiple results were submitted for \textit{FreeCAD} (2 sets), \textit{FZKViewer} (2 sets),\textit{ Autodesk Revit} (6 sets, multiple versions), \textit{Vectorworks Designer} (2 sets), \textit{Archicad} (3 sets).
These offer the opportunity for timing comparisons to investigate the impact of hardware on software performance. 
The graph in \Reffig{fig:successrates} gives a summary of the success rates returned for the tests on the three datasets.
The graph in \Reffig{fig:timinggraph} gives the count of the different timing values for the successful tests.

Note that in some cases users reported results for some of the tests but not for all of them (``No result reported''). Additionally, some users typed in comments such as ``no error'' instead of giving specific timing. These are included in the ``No result reported'' count.

For the \textit{Myran.ifc} dataset, none of the tasks took more than 5 minutes to execute, and the large majority of the software packages tested could zoom and pan the data immediately.
Data loading times of less than a minute were also dominant (65\% in total). 28 out of the 43 participants reported successful IFC export, with 86\% of these exports being executed in less than a minute.

The impact of the more complex dataset (adopting IFC v.4) on the tests is visible in the results, with only 58\% of the software packages able to carry out the basic timed test that involved opening the dataset in the software, and 35\% reporting that the software crashed.
For the \textit{Savigliano.ifc} dataset, in contrast to the \textit{Myran.ifc} dataset, 9 of the successful attempts to open the data (36\%) took 5 or more minutes to execute, although once opened zooming and panning the data was immediate, and the large majority of the software packages tested could zoom and pan the data immediately.
Data loading times of less than a minute were also dominant (65\% in total).
16 out of the 43 participants reported successful IFC export, with all of these taking less than 5 minutes to execute.

The results with the \textit{UpTown.ifc} data demonstrate clearly the impact of the larger dataset on the tests carried out, with only 47\% of the software packages able to carry out the basic timed test that involved opening the dataset in the software, and 41\% reporting that the software crashed.
In addition, in contrast to the \textit{Myran.ifc} dataset, 9 of the 20 successful attempts to open the data (55\%) took 5 or more minutes to execute, with 2 of the participants reporting test times of between 20 minutes and 1 hour, and 3 of over one hour.
Interestingly, once open, the vast majority of the testers reported sub-minute execution times for zooming and panning.
Only 15 out of the 43 participants reported successful IFC export, with more than 50\% of them taking over 20 minutes.

\subsubsection{Multiple Tests on Same Software Packages}

As noted above, the crowdsourced approach taken in this project resulted in multiple participants testing the same software, providing the opportunity for comparison.
For both \textit{FreeCAD} and \textit{FZK viewer}, one out of the two respondents reported that they were unable to open the \textit{Myran.ifc} dataset.
\textit{Vectorworks} results are similarly consistent, with the only minor difference being that one participant reported that the \textit{Myran.ifc} data opened immediately and another that it took less than one minute (both on machines with 16GB RAM). 

Identical timing results are reported for the zoom/pan/rotate/query tasks for the remaining three packages (\textit{Autodesk, ArchiCAD}), and as noted above, the large majority of software packages managed these functions in near immediate time.
As different analytical tasks were selected by participants these are not comparable. 

Differences do, however, arise when execution time for the remaining functionality is considered.
\Reftab{tab:multipletiming} summarises the remaining timing results for these packages for the \textit{Myran.ifc} dataset, which was chosen as this proved the least problematic of the three datasets in terms of successful execution of tasks.

\begin{table}[htbp]
\small
\centering
\begin{tabular}{|p{4cm}|p{3cm}|p{4cm}|p{4cm}|}
\hline
& \textbf{Import} & \textbf{Inspect Linked Objects} & \textbf{Export to IFC} \\
\hline
\textit{Autodesk Revit 2019 --- Result 1} & 1--5 minutes & it's almost immediate & it's almost immediate \\
\hline
\textit{Autodesk Revit 2018 --- Result 2} & 1--5 minutes & it's almost immediate & 1--5 minutes \\
\hline
\textit{Autodesk Revit 2018 --- Result 3} & 1--5 minutes & the software does not allow this & 1--5 minutes \\
\hline
\textit{Autodesk Revit 2018 --- Result 4} & 1--5 minutes & the software does not allow this & crashes \\
\hline
\textit{Autodesk Revit 2019.2 --- Result 5} & 1--5 minutes & the software does not allow this & it's almost immediate \\
\hline
\textit{Autodesk Revit 2020 --- Result 6} & less than a minute & less than a minute & it's almost immediate \\
\hline
\textit{ArchiCAD --- Result 1} & less than a minute & the software does not allow this & less than a minute \\
\hline
\textit{ArchiCAD --- Result 2} & less than a minute & the software does not allow this & less than a minute \\
\hline
\textit{ArchiCAD --- Result 3} & 1--5 minutes & it's almost immediate & less than a minute \\
\hline
\end{tabular}
\caption{Inconsistent timing results that were submitted for the \textit{Myran.ifc} dataset}%
\label{tab:multipletiming}
\end{table}

For \textit{Revit}, there is an apparent performance improvement when opening \textit{Myran.ifc} data in the 2020 version of the software, and an improvement of export time from 1--5 minutes in the 2018 version, through to ``almost immediate'' in the 2019 and 2020 versions of the software.
In terms of hardware, all tests were carried out on Windows 10 machines.
However, the machine used for \textit{Revit 2020} had 64GB of RAM, with Result 1 obtained with 16GB RAM, and Result 2, 4 and 5 with 8GB RAM\@.
Contradictory results in the linked object query make these difficult to compare. 
For \textit{ArchiCAD}, all users report an IFC export time for \textit{Myran.ifc} of less than a minute, but two report that it was not possible to query linked information.
Examining the hardware used, Result 3 for \textit{ArchiCAD} –-- file opening time of 1--5 minutes --– was obtained on a machine with 8GB RAM, whereas the others were tested on machines with 16GB RAM (all Windows 10 machines, all with dedicated graphics cards).

\section{Discussion}\label{sec:discussion}

What is apparent while analysing the results of the exported files, is that it is very difficult to find a pattern or a sensible explanation about what happened in the export phase, since there are very few cases in which the IFC entities are completely kept / lost / gained in all the exported models.
The most frequent scheme is: some models keep all --- some models lose all --- some models keep part --- some entities are added. The very high complexity of both the semantic data model and the geometry management is for sure related to the issue.

Similarly, the investigation about loading IFC files suggests a general difficulty in reading semantics correctly, very little support for georeferencing and quite good support for general geometry, as visualized, which is not unambiguously read though, when looking at the IFC geometries datasets results.
But no clear pattern is found, that could make it possible to better understand the specific issues of IFC interoperability.
A clear result is just that little interoperability is actually reached, since there are very few tools able to read the standardised datasets correctly and even fewer that are able to export them consistently.
The ability to uniquely interpret the models and to leave them consistent through the import-export phases is absolutely essential for interoperability and what it enables (data exchange, data re-use and so on).
At this stage it is not possible to fully trust standardised models though, even just for simple file exchange.

Results were reported for 33 software packages, highlighting the wide range of tools available to users.
Different versions of \textit{Revit} also demonstrated that \textit{Autodesk} is making improvements to the interoperability of the software --- \textit{Revit 2019} gave results for all data, but \textit{Revit 2018} could not handle two of the three datasets (\textit{Uptown.ifc} and \textit{Savigliano.ifc}).
\textit{Myran.ifc} could be opened by 40 participants, \textit{Savigliano.ifc} by 25 and \textit{Uptown.ifc} by 20.
The twelve participants who reported that they were not able to open \textit{Uptown.ifc} also reported the same issue with \textit{Savigliano.ifc}.
Ten participants reported problems with \textit{Uptown.ifc} but none with \textit{Savigliano.ifc} and 4 with \textit{Savigliano.ifc} but not with \textit{Uptown.ifc}, indicating that the size of the dataset is, overall, more problematic than the IFC version.

Regarding performance, the impact of the size of the dataset on overall performance is marked – with 55\% of the software packages that managed to open the \textit{Uptown.ifc} dataset reporting a time of 5 minutes or more. 
While it is not possible to say which software package is the fastest --– the approach to timing used general timing categories rather than requiring the user to undertake the onerous task of time measurement --- and performance will also depend both on the hardware being used, we can report that none of the software packages managed to carry out all the visualisation tasks in under a minute for the \textit{Uptown.ifc} dataset, although 9 packages achieved this with the \textit{Myran.ifc} dataset and 9 with the \textit{Savigliano.ifc} dataset.
For IFC export, 24 software packages managed to export the data in 1 minute or less for \textit{Myran.ifc}, 14 for \textit{Savigliano.ifc} but only two for \textit{Uptown.ifc}.


The set of available functionalities of software to work with IFC models is moreover quite limited.
The analysis about the model itself (Type 1) are essential to make sure a model is valid and suitable to support further analysis.
On the other hand, tools are necessary to use IFC models for the operational intentions it was conceived for.
However, not many tools are now able to work with IFC effectively.

%
%

The fact that the inquiry is based on voluntary and completely open contribution can be considered both as a strong point and a limitation of this work.
On the one hand, it is essential to cover the investigated object in the most thorough way: as many software packages as possible, with many experts involved, but also with the inclusion of less expert users to test also user-friendliness.
However, one outcome of this approach is some incompleteness in the tools that are assessed\footnote{\eg\ the open source \textit{xBIM} (\url{https://docs.xbim.net/index.html}), the very recent BlenderBIM (\url{https://blenderbim.org/}) and some of the niche engineering tools were not tested.}.
This one was limited by the ex-post integration of tests about the software not originally considered.
Moreover, the tests reporting suspicious results according to the promoting team experiences (for both too good or too bad performances), were double-checked with new tests or by asking for clarifications.
A further issue could be the inexperience of some testers, reporting about tools behaviour in an inaccurate way.
To lower this eventuality, it was checked that all the delivered answers were described with sufficient care, whatever the level of expertise could be. Once verified this, eventual conflicting answers with respect to the tools actual potential, could indicate a deficiency in the suitability of the tool to be used by any inexpert user (which would be anyhow necessary for the models to be used in practice).

The involvement of a large part of the community (twice the number of participants delivering the results initially registered to the initiative) is also important to perceive the relevance of the topic, although dealing with a somehow hidden issue among the high-level standardization and academic communities.

\subsection{Lessons learnt and consequent suggestions}

The insight and data gained with this study can be the base for reasoning about improvements to the IFC model, the common behaviour of the models and the most urgent issues to be tackled by software, for example, the suitable implementation of georeferencing.

Control in export and import would be critical to make the choice explicit about the used possible alternative to store information in IFC\@.
Although possible ambiguities and equivalent alternatives should be avoided as much as possible, in some cases, applications could require information to be stored according to specific structures, considering the slight differences between them (as an example, see the difference between LoGeoRef30 and LoGeoRef40, or the generalization level of semantics required by applications).

Such control, for georeferencing would consist in the choice about the LoGeoRef to be read, when alternative ones are stored (including LoGeoRef50 enabled by IFC4).
It could be both a choice, to be selected during import, and an instruction, as a requirement for the model to be input.
A reasoning could be made about the opportunity of allowing the reading of N-E, H and rotation separately, according to different LoGeoRefs.

The georeferencing issue is even more articulated.
First, regardless of the country and its level of advancement in implementing the BIM idea in ongoing projects, the little use by designers and architects of BIM models with the right georeference is similar.
Almost all designers expect a model in their local coordinate system, despite the fact that the point cloud obtained for this purpose has the correct coordinates, consistent with the national system.
To make it worse, in most cases, ``long'' coordinates are truncated due to software problems with ``long'' coordinates.
Therefore, in addition to developing effective methods for storing georeferencing in IFC files, it seems necessary to provide education to designers and architects and a clear guideline about how to deal with it.

With respect to semantics, probably a smaller set of entities, to be used according to stricter rules would help in implementing a suitable interpretation in software.
At least BIM software, supposed to have the most general scope, should try to implement internal structures of entities/ families in the native format in agreement with the IFC data model, as much as possible.
This would be not supposed to be a top-down directive from buildingSMART, but probably an agreement coming from the collaboration of software developers with standardization institution, starting from the needs of practitioners and users.

If such a stricter collaboration would be implemented, probably also the IFC model could have an indication for selecting its most important core part to be maintained, improved and likely simplified or made less ambiguous.
The parts of the model more specific of different application domains could be cared together with the specific expert domains-practitioners-developers.

Control and transparency would be the second requirements as well.
It would be an added value in the case of export: How will the software export the data? For example, a software-specific but explicit MVD could describe exactly how the internal model of the software is mapped on which part of the IFC model.
And it would help the import phase as well: how will be the IFC classes  interpreted? Which ones are supposed to be interpreted exactly, since part of the software data model? Which ones could instead be misinterpreted and how: maybe a choice could be provided to map the un-recognized classes, or they could be represented as their parent class, or as a generic class. This should be a straightforward, guided and clear process instead of an advanced option of the software likely supposed to be used mainly by experts.

Since it is not possible to detect any pattern in the behaviour of software, the issue is more identifiable in such indefiniteness than in one specific aspect of the standard or of the software.

One unique way of writing the IFC files should also be imposed, to avoid redundancy or ambiguity in the storage of information.
Semantics would have advantages from simpler and clearer rules about how to structure entities, priorities and limits. For implementations, and therefore practice, it is for sure preferable to rely on a simpler model than a very complex and expressive but inconsistent one.


Finally, reducing the set of geometries in the schema, together with clear constraints on how and when to use these geometries would enable software developers to implement them more robustly, avoiding loss of information, intersections and so on. As is the aim already in future versions of IFC, this should also be made more use-case dependent. When an export is intended to function as a static reference for visualization, coordination or inference checking, procedural and parametric representation should be largely avoided. When transfer of design intent and parametric behavior is required the more procedural geometry definitions can be used.
In addition, the non-compliances possibly implemented by some software, should at least give an error message and make the user explicitly choose if using a non-compliant IFC file anyway, to the detriment of interoperability. This should be part of the IFC certification process too. One potential avenue to detect flaws in the geometric interpretation is to serialize a set of descriptive properties (such as center of mass, surface area, volume, surface genus) along with the geometry during export that the importing application can calculate from its reinterpreted geometry representation and compare, so called geometric validation properties from \citet{PRATT20051251} in the case of STEP\@.

Again, implementing some control on the export of geometries by modelling software would help in making the (likely few) possible choices explicit, fostering re-usability and interoperability. The storage of such choices in metadata would also be useful.

As a general rule, the use of default settings while importing or exporting files to overcome the presence of undefined information should always be avoided, since it is source of many misleading inconsistencies (\eg\ inferring CRS).

In some case, it could be possible to object that the IFC aim is different from some of the aspects tested here.
For example, someone could say that editing is not a priority or not even a planned possibility for IFC files, that should only be exchanged as reference:
\eg\ for different professionals designing the same building, for which semantics are barely needed; only for analysis (asset and facility management, project management, project analysis, computations and so on), where semantics and quality do matter.
If this is the case, the purpose of IFC itself should be clearly stated and maybe reduced.

\section{Conclusions}\label{sec:conclusion}

The described study in this paper, part of the GeoBIM benchmark project, was intended to point out and provide evidence about the support for and issues of available software for standardised information in IFC\@.
Interoperability is essential for a number of use cases, and even for simply exchanging and re-using data.
Standards are supposed to be enabling such interoperability, and standardization is the essential premise to the development of any integration, including the GeoBIM one.

The Industry Foundation Classes, by buildingSMART, is the official reference open standard for BIM management and storage.
However, we have personally stumbled across a number of issues while using IFC, and many more have been often informally reported to us by practitioners.
These prevent the effective use of IFC in practice, without the issues being clear to users, who usually rely on proprietary formats (such as the \textit{Revit} native format), using them as often more reliable de-facto standards.
Moreover, although we were aware that some data or functionality loss was prevalent when using IFC, no systematic proof was available in the literature.
This project aimed to create such proof, which could be used as the base of future improvements in implementations, in data modelling and in the standard itself.

The results show how difficult it is for software tools to read, manage and export the IFC datasets consistently, in a manner that woud take full disadvantage of interoperability, and that would enable the effective usability of datasets in use cases too.

Besides the internal interoperability of IFC in BIM software, the potential integration with 3D geoinformation and respective standards is additionally hindered.
It is clear that if even IFC software cannot consistently read and write IFC, multi-format software will do even worse, especially when dealing with the areas where both domains diverge (\eg\ georeferencing and semantics).
In fact, it is almost impossible to plan, design and implement effective solutions for mapping, conversions and object transformation to other formats without relying on a stable and trustworthy initial format. 

We are aware that there is a possible bias in this study's results, which could be produced by the lack of expertise of some participants, or by the initial inaccuracies in the provided datasets.
However, such a bias would also reflect the way in which users use software and product data in practice.
Nevertheless, in order to minimise potential bias, the datasets were validated and improved as much as possible for the purpose of the benchmark, in order to limit their influence on the quality and reliability of results.
Therefore, if there are still problems in the datasets, it would likely reflect additional drawbacks of the standard itself, for the little clarity about its use for the modelling of actual datasets and the difficulty in implementation, which could produce little intuitive tools.

This study shows the drawbacks of the current IFC standard and implementation, and the related difficulties, also due to the challenges to which it is intended to respond (\eg\ representation of the information regarding a vast and complex field, flexibility to multiple needs).
A particularly important point is that the outcomes of the study are officially documented here, in a manner that they can be the base for future research in the field and development of concrete solutions, such as the addition of constraints and specific guidelines, simpler ways to store geometry, better selection of useful semantics, and so on.

Giving clear guidance towards changes in the IFC schema is out of scope in the present paper.
However, it is clear that from the point of view of software support, more consistent implementations of the existing schema are needed.
There are many possibilities that could help in this regard, \eg\ the creation of high quality open source libraries to read/write IFC, the creation of reference implementations, and the writing of more extensive documentation that shows how the IFC data model is mapped to software's internal data models.
All of these would enable the reuse of source code and the definition of best practices.

Considering the results of this study as evidence, future work should be aimed at filling in gaps in the results (\eg\ through the study of additional software or new reference datasets), updating the results with respect to new software, and the solution of the outlined issues.

For example, a clear area of opportunity is the definition of clear guidelines about the use of complex semantics, fixing priorities and criteria for their use.
Moreover, specific kinds of geometries should be selected for specific cases, adding constraints to guarantee that software can import, read, use and re-export them without any change.
In addition, when considering the performance related to the computational requirements, we can easily understand that the reduction of data size is urgent.

buildingSMART promisingly began working towards solving such issues, trying to work for less complex models offering more straightforward choices, easier to implement (see the ``Ten principles for the future IFC'' by buildingSMART\footnote{\url{https://github.com/buildingSMART/NextGen-IFC/wiki/Ten-principles-for-a-future-IFC}}) and in the work towards IFC v.5. 
Thanks to this study, it was possible to gain a higher awareness of the specific issues to be tackled in future research in order to foster an improved adoption of the IFC standard.

\section{ACKNOWLEDGEMENTS}

This work was possible thanks to the collaboration of the whole GeoBIM benchmark team (with their work as in-kind contribution to the project), all the data providers and the participants making the tests, listed in the GeoBIM benchmark website\footnote{\url{https://3d.bk.tudelft.nl/projects/geobim-benchmark/participants.html}}.

The benchmark was funded by ISPRS and EuroSDR\@. This project has also received funding from the European Research Council (ERC) under the European Union's Horizon 2020 Research \& Innovation Programme (grant agreement no. 677312, Urban modelling in higher dimensions) and Marie Skłodowska-Curie (grant agreement No. 707404, Multisource Spatial data Integration for smart City Applications).


\bibliographystyle{plainnat}
\bibliography{T1-finalbenchmark-rev}

\end{document}